\documentclass[11pt,twoside,nohyper]{jhep}
\usepackage{epsfig}
\usepackage{latexsym}
\usepackage{euscript}
\usepackage{amssymb}
\usepackage{fancyheda}
\usepackage{times}
\usepackage[varg]{txfonts}
\usepackage{caption}
\newcommand{\ftn}{\footnotesize}

\newcommand{\ssz}{\scriptsize}

\newcommand{\TeV}{{\mbox{\rm TeV}}}

\newcommand{\GeV}{{\mbox{\rm GeV}}}
\newcommand{\keV}{{\mbox{\rm keV}}}
\newcommand{\vH}{{\mbox{$\bar H$}}}
\newcommand{\vHi}{{\mbox{$\bar H_{{\rm I}}$}}}
\newcommand{\vrho}{{\mbox{$\bar\rho$}}}
\newcommand{\vn}{{\mbox{$\bar n$}}}
\newcommand{\vq}{{\mbox{$\bar q$}}}
\newcommand{\vs}{{\mbox{$\bar s$}}}
\newcommand{\vqi}{{\mbox{$\bar q_{\rm I}$}}}
\newcommand{\vQ}{{\mbox{$\bar Q$}}}
\newcommand{\vV}{{\mbox{$\bar V$}}}
\newcommand{\vVo}{{\ensuremath{\bar V_0}}}

\newcommand{\vTi}{{\ensuremath{T_{{\rm I}}}}}
\newcommand{\vti}{{\ensuremath{\vtau_{{\rm I}}}}}
\newcommand{\vtif}{{\ensuremath{\vtauf_{_{\rm I}}}}}
\newcommand{\vtf}{{\ensuremath{\vtau_{{\rm F}}}}}
\newcommand{\vtns}{{\ensuremath{\vtau_{{\rm BBN}}}}}
\newcommand{\vtkr}{{\ensuremath{\vtau_{{\rm KR}}}}}
\newcommand{\vtp}{{\ensuremath{\vtau_{{\rm ext}}}}}
\newcommand{\vtx}{{\ensuremath{\vtau_{{\chi}}}}}
\newcommand{\vts}{{\ensuremath{\vtau_{{\rm SUSY}}}}}
\newcommand\cqg[3]{{\sl Class.\ Quant.\ Grav.\ }{\bf #1}, #3 (#2)}

\def\openep{\leavevmode\hbox{\normalsize$\iota$\kern-3.8pt$^$-}}
\newcommand{\vtauf}{\ensuremath{\tauup}}
\newcommand{\vtau}{\ensuremath{\tauup}}

\def\bec{\begin{center}}
\def\eec{\end{center}}
\def\beq{\begin{equation}}
\def\eeq{\end{equation}}
\def\bea{\begin{eqnarray}}
\def\eea{\end{eqnarray}}

\newcommand{\Ti}{\ensuremath{T_{\rm I}}}
\newcommand{\Tkr}{\ensuremath{T_{\rm KR}}}
\newcommand{\Ts}{\ensuremath{T_{\rm SUSY}}}
\newcommand{\Tns}{\ensuremath{T_{\rm BBN}}}
\newcommand{\Tc}{\ensuremath{T_{\rm C}}}

\newcommand{\OmX}{{\mbox{$\Omega_X h^2$}}}
\newcommand{\Omax}{{\mbox{$\Omega_{\ax} h^2$}}}
\newcommand{\Omgr}{{\mbox{$\Omega_{\Gr} h^2$}}}
\newcommand{\Omx}{{\mbox{$\Omega_\chi h^2$}}}
\newcommand{\Domx}{\ensuremath{\Delta\Omega_{\chi}}}
\newcommand{\Ygr}{\ensuremath{Y_{\Gr}}}

\newcommand{\Omqns}{\ensuremath{\Omega_q^{\rm BBN}}}

\newcommand{\vrhoA}{{\ensuremath{\bar\rho_{{\rm A}}}}}
\newcommand{\vrhoM}{{\mbox{$\bar\rho_{{\rm M}}$}}}
\newcommand{\vrhoB}{{\ensuremath{\bar\rho_{{\rm B}}}}}
\newcommand{\vrhoR}{{\ensuremath{\bar\rho_{{\rm R}}}}}

\newcommand{\vsv}{\ensuremath{\overline{\sgv}}}
\newcommand\sgv{\langle \sigma v\rangle}

\newcommand\Gm[1]{\Gamma_{#1}}

\newcommand{\gl}{\ensuremath{\tilde{g}}}
\newcommand{\ax}{\ensuremath{\tilde{a}}}
\newcommand{\sq}{\ensuremath{\tilde{q}}}
\newcommand{\Gr}{\ensuremath{\widetilde{G}}}
\newcommand{\nt}{\ensuremath{\chi}}
\newcommand{\nequ}{\ensuremath{n^{\rm eq}}}
\newcommand{\ps}{\ensuremath{e^+}}
\newcommand{\el}{\ensuremath{e^-}}

\newcommand{\mx}{{\mbox{$m_\chi$}}}
\newcommand{\mxx}{{\mbox{$m_{\ax}$}}}
\newcommand{\mgr}{{\mbox{$m_{\Gr}$}}}

\def\ewimps{\emph{e}-WIMPs~}

\newcommand{\sFref}[2]{Fig.~\ref{#1}-{\small\sf ({#2})}}
\newcommand{\sEref}[2]{Eq.~(\ref{#1}{\small\sf {#2}})}
\newcommand{\eref}[1]{(\ref{#1})}

\newcommand{\Eref}[1]{Eq.~(\ref{#1})}
\newcommand{\Sref}[1]{Sec.~\ref{#1}}
\newcommand{\Fref}[1]{Fig.~\ref{#1}}
\newcommand{\Tref}[1]{Table~\ref{#1}}
\newcommand{\cref}[1]{Ref.~\cite{#1}}
\newcommand{\etal}{{\it et al.\/}}

\newcommand{\mP}{\ensuremath{m_{\rm P}}}

\newcommand\eqs[2]{Eqs.~(\ref{#1}) and (\ref{#2})}

\newcommand\ssFref[3]{Fig.~\ref{#1}-{\small\sf ({#2})} and
{\small\sf ({#3})}}

\title{\huge \bfseries\scshape Tracking Quintessence and\\ Cold Dark Matter Candidates}

\author{\large \bfseries\scshape S. Lola, C. Pallis and E. Tzelati\\
Department of Physics, University of Patras,
\\ Panepistimioupolis, GR-265 00 Patras, GREECE \\ \vspace{5pt}
}

\abstract{We study the generation of a kination-dominated phase in
the context of a quintessential model with an inverse-power-law
potential and a Hubble-induced mass term for the quintessence
field. The presence of kination is associated with an oscillating
evolution of the quintessence field and the barotropic index. We
find that, in sizeable regions of the parameter space, a tracker
scaling solution can be reached sufficiently early to alleviate
the coincidence problem. Other observational constraints
originating from nucleosynthesis, the inflationary scale, the
present acceleration of the universe and the dark-energy-density
parameter can be also met. The impact of this modified
kination-dominated phase on the thermal abundance of cold dark
matter candidates is investigated too. We find that: (i) the
enhancement of the relic abundance of the WIMPs with respect to
the standard paradigm, crucially depends on the hierarchy between
the freeze-out temperature and the temperature at which the
extrema in the evolution of the quintessence field are
encountered, and (ii) the relic abundance of \ewimps takes its
present value close to the temperature at which the earliest
extremum of the evolution of the quintessence field occurs and, as
a consequence, both gravitinos and axinos arise as natural cold
dark matter candidates. In the case of unstable gravitinos, the
gravitino constraint can be satisfied for values of the initial
temperature well above those required in the standard cosmology.

{\sl\bfseries Published in} {\sl  J.
Cosmol. Astropart. Phys.} {\bf 11}, 017 (2009)}

\begin{document}

\setcounter{page}{1} \pagestyle{fancyplain}

\addtolength{\headheight}{.5cm}

\rhead[\fancyplain{}{ \bf \thepage}]{\fancyplain{}{\scshape
Tracking Quintessence and CDM Candidates}}
\lhead[\fancyplain{}{\sc \leftmark}]{\fancyplain{}{\bf \thepage}}
\cfoot{}

\section{Introduction}\label{intro}

A plethora of recent data \cite{wmap, snae} indicates
\cite{wmapl} that the two major components of the present universe
are \emph{Cold Dark Matter} (CDM) and \emph{Dark Energy}
(DE) with density parameters \cite{wmap}:
\beq \mbox{\sf\small (a)}~~\Omega_{\rm
CDM}=0.214\pm0.027~~\mbox{and}~~\mbox{\small\sf (b)}~~\Omega_{\rm
DE}=0.742\pm0.03, \label{cdmba}\eeq
at $95\%$ \emph{confidence level} (c.l.). Identifying the nature
of  these two unknown substances is one of the major challenges
in modern cosmo-particle theories.

Among the natural candidates \cite{candidates} to account for CDM
are \cite{jungman} the \emph{weakly interacting massive particles}
(WIMPs) \cite{lkk, wimps}, with prominent representative in
supersymmetric (SUSY) theories the lightest neutralino
\cite{goldberg} and the \emph{extremely WIMPs} (\emph{e}-WIMPs)
\cite{ewimps} with most popular representatives, the gravitino,
$\Gr$, and the axino, $\ax$. Assuming $R$-parity conservation, the
\emph{lightest SUSY particle} (LSP) is stable and can be either a
WIMP or an \emph{e}-WIMP, in a sizeable region of the SUSY
parameter space. The interactions of WIMPs ensure that they come
to chemical equilibrium with the plasma and decouple from it at a
temperature $T_{\rm F}\sim (10-20) ~{\rm GeV}$. On the other hand,
the interaction of \emph{e}-WIMPs  are \emph{extremely} weak since
they are suppressed by the reduced Planck scale, $m_{\rm P}=M_{\rm
P}/\sqrt{8\pi}$ (where $M_{\rm P}=1.22\cdot10^{19}~{\rm GeV}$ is
the Planck mass) in the case of $\Gr$ and by the axion decay
constant, $f_a\sim(10^{10}-10^{12})~{\rm GeV}$ (for a review, see
Ref.~\cite{kim}) in the case of $\tilde a$. Consequently,
\emph{e}-WIMPs depart from chemical equilibrium very early and
their relic density (created due to this early decoupling) is
diluted by primordial inflation. Subsequently, \emph{e}-WIMPS can
be reproduced in the thermal bath through scatterings
\cite{gravitinoc, Bolz, axino, steffenaxino} and decays
\cite{axino, small, strumia} involving superpartners. For both
WIMPs and \emph{e}-WIMPs (hereafter denoted collectively as $X$)
we may have extra non-thermal contributions to their relic
density, $\Omega_X h^2$, \cite{gravitinont, axinont}, from the
out-of-equilibrium decay of the \emph{next-to-LSP} (NLSP).
However, this mechanism is highly model dependent since it is
sensitive to the nature of the NLSP, its decay products and the
extra restrictions which have to be imposed in order to maintain
the success of the standard \emph{Big-Bang Nucleosynthensis}
(BBN). Given that it does not modify the results in any essential
way, other than by moderate factors, we opt to keep the analysis
as generic as possible, and therefore focus on the thermal
production of CDM candidates.

A particle $X$ consists a viable CDM candidate, provided its relic
density $\Omega_X h^2$ can be confined in the region \cite{wmap}
\beq\label{cdmb} \mbox{\sf\small (a)}~0.097\lesssim \Omega_X h^2
\lesssim 0.12~~\mbox{for}~~\mbox{\sf\small (b)}~10~\keV<m_X<m_{\rm
NLSP} \eeq
where $m_{\rm NLSP}$ is the NLSP mass and the lower bound on $m_X$
comes from the fact that lower $m_{X}$'s cannot explain
\cite{sformation} the observed  early reionization \cite{wmap}.
Clearly, the $\Omega_X h^2$ calculation crucially depends on the
assumption on the dominant component of the universe during the
decoupling of WIMPs or the reproduction of \emph{e}-WIMPs. In the
\emph{standard cosmological scenario} (SC) we assume that
primordial inflation is followed by the \emph{radiation dominated}
(RD) epoch. However, our current knowledge of the history of the
universe before BBN, also allows for other possibilities (see,
e.g., \cite{Kam, several, masiero}). Among them, an interesting
alternative is provided by the presence of a \emph{kination
dominated} (KD) \cite{kination} post-inflationary era, which
enhances the thermal abundance of WIMPs \cite{Kam, salati, prof,
jcapa} and significantly reduces thermal abundance of
\emph{e}-WIMPs \cite{huelva} \emph{with respect to} (w.r.t) their
values in the SC.

The existence of a KD era is an open possibility in the framework
of \emph{quintessential scenaria} (QS). Quintessence \cite{early}
(for reviews, see Ref.~\cite{der}) is a scalar field, slowing
evolving today, which can provide the required amount of the
present vacuum energy and therefore explain (at least at the
classical level) the other major component of the universe, the
DE. Kination is also an indispensable ingredient of the
quintessential inflationary scenaria \cite{qinf, dimopoulos,
chung}. In recent papers \cite{jcapa, huelva} we considered the
creation of a KD era in the context of the exponential
quintessential model \cite{wet, expo}, taking into account a
number of relevant phenomenological requirements. Although this
model can reproduce a viable present-day cosmology in conjunction
with the domination of an early KD era, for a reasonable region of
initial conditions \cite{brazil}, it does not possess a
tracker-type solution \cite{salati, attr}, where quintessence is
able to reach the required value today starting from a very wide
set of initial conditions in the remote past. This attractive
behavior occurs in models with inverse power-law potentials, which
are naturally expected in high energy particle physics models
\cite{binetruy}. In this way the so-called ``coincidence'' or
``why now'' problem, related to the fact that the quintessential
energy density is such that it is dominating the cosmic expansion
right now, is addressed. However, within the minimal realizations
of these models, these positive features do not coexist with the
presence of an early KD era \cite{attr, mas}.

In the current work, we reconsider the generation of an early KD
era in the context of tracking quintessence, switching on a
Hubble-induced time dependent mass term for the quintessence
field, along the lines of \cref{mas} (see also \cref{Hq, davis,
Tq}). The paper is structured as follows: In Sec.~\ref{sec:quint},
we show that a KD era within this framework, is characterized by
an oscillatory evolution of the quintessence field and the
barotropic index. Observationally acceptable values for the latter
at present times can be obtained for relatively law values  of the
exponent in the potential \cite{bacci}. Other restrictions arising
from BBN, the inflationary scale and the DE density parameter can
also be met in a wide range of the parametric space, which turns
out to have a band structure. In Sec.~\ref{sec:boltz}, we
investigate the impact of this KD era on the thermal abundance of
WIMPs and \emph{e}-WIMPs, solving both numerically and
semi-analytically the relevant Boltzmann equations. We find that,
if $X$ is a WIMP, $\OmX$ depends crucially on the hierarchy
between the freeze-out temperature and the temperature at which
the extrema in the evolution of the quintessence field are
encountered. On the other hand, if $X$ is an \emph{e}-WIMP, $\OmX$
is determined mainly at the temperature where the first extremum
(after the onset of KD era) in the evolution of the quintessence
field occurs. In \Sref{sec:wimp} and \ref{sec:ewimp} respectively,
we study the dependence of $\OmX$ on the free parameters of the
theory, and identify the allowed parameter space for WIMPs and
\emph{e}-WIMPs. Our conclusions are summarized in
Sec.~\ref{sec:con}. For completeness, we also discuss the
cosmology of unstable $\Gr$ within our QS in Appendix A.

Throughout the text, brackets are used by applying disjunctive
correspondence. Natural units are assumed for the Planck's and
Boltzmann's constants and for the velocity of light
($\hbar=c=k_{\rm B}=1$). The subscript or superscript ``$0$''
refers to present-day values (except in the coefficient $\vVo$)
and $\log~[\ln]$ stands for logarithm with basis $10~[e]$.
Finally, we assume that the domain wall number \cite{kim} is equal
to 1.

\section{Tracking Quintessence}\label{sec:quint}

In this section we outline the several aspects of tracking
quintessence (Sec.~\ref{Beqs}) and the various observational
restrictions that have to be imposed (Sec.~\ref{reqq}). We then
highlight the scalar field dynamics in \Sref{Qev} and describe the
allowed parameter space of our QS in Sec.~\ref{ap}.

\subsection{The Quintessential Set-up}
\label{Beqs}

We present below the equations which govern the evolution of the
quintessence field (\Sref{Beqs1}) and the method we use in order
to solve them numerically (\Sref{Beqs2}).

\subsubsection{Relevant Equations.\label{Beqs1}}
We assume the existence of a spatially homogeneous scalar field
$q$ (not to be confused with the deceleration parameter
\cite{wmapl}) which obeys the equation:
\beq \ddot q+3H\dot
q+V_{,q}=0,~~\mbox{where}~~V=V_a+V_b~~\mbox{with}~~V_a={M^{4+a}\over
q^a}~~\mbox{and}~~V_b={b\over2}H^2q^2,\label{qeq} \eeq
is the adopted potential for the field $q$ with $M$  a mass scale
and $,q$ [dot] stands for the derivative w.r.t  $q$ [the cosmic
time, $t$]. The main feature of $V_a$ is the existence of
'tracker' solutions, which are attractors \cite{attr, trak} in
field space, while $V_b$ (with $b$ of order unity) expresses a
quite generic interaction term which arises e.g. due to
non-canonical terms of the K\"ahler potential of $q$ \cite{Hq,
davis}. Similar interactions \cite{Tq} arise due to the thermal
effects as well. As shown in \cref{mas} and verified in \Sref{ap},
a mild tuning of the coefficient $b$ enlarges the configuration
space that leads to the desirable insensitivity to the initial
conditions, without modifying the behavior of the field today.

The Hubble expansion parameter $H$ in \Eref{qeq} is given by
\begin{equation}\label{rhoqi}
H= \sqrt{\rho_q +\rho_{{\rm R}}+ \rho_{{\rm M}}}/\sqrt{3}m_{{\rm
P}}~~\mbox{with}~~\rho_q=\frac{1}{2}\dot q^2+V,\eeq
the energy density of $q$. The energy density of radiation,
$\rho_{{\rm R}}$, can be evaluated as a function of temperature,
$T$, whilst the energy density of matter, $\rho_{{\rm M}}$, with
reference to its present-day value:
\beq\label{rhos} \rho_{{\rm R}}=\frac{\pi^2}{30}g_{\rho*}\
T^4~~\mbox{and}~~\rho_{{\rm M}}R^3=\rho_{{\rm M0}}R_0^3
\end{equation}
with $R$ the scale factor of the universe. Assuming no entropy
production due to the domination of $q$ or another field, the
entropy density, $s$, satisfies the equation
\beq sR^3=s_0R_0^3~~\mbox{where}~~s=\frac{2\pi^2}{45}g_{s*}\ T^3,
\label{rs}\eeq
where $g_{\rho*}(T)~[g_{s*}(T)]$ is the energy [entropy] effective
number of degrees of freedom at temperature $T$. Their precise
numerical values are evaluated by using the tables included in
public packages \cite{micro}, assuming the particle spectrum of
the Minimal SUSY Standard Model.

\subsubsection{Numerical Integration. \label{Beqs2}} The numerical integration
of Eq.~(\ref{qeq}) is facilitated by converting the time
derivatives to derivatives w.r.t the logarithmic time
\cite{brazil} which is defined as a function of the redshift $z$:
\beq \vtau=\ln\left(R/R_0\right)=-\ln (1+z).\label{dtau} \eeq
Changing the differentiation and introducing the following
dimensionless quantities:
\beq \label{vrhos}\vrho_{{\rm R}}=\rho_{{\rm R}}/\rho_{\rm
c0},~\vrho_{{\rm M}}=\rho_{{\rm M}}/\rho_{\rm
c0},~\vV_a=V_a/\rho_{\rm c0},~\vH={H/
H_0},~~\mbox{and}~~\vq=q/\sqrt{3}m_{{\rm P}},\eeq
Eq.~(\ref{qeq}) turns out to be equivalent to the system of two
first-order equations:
\bea &\vQ=\vH\vq^\prime
~~&\mbox{and}~~\vH\vQ^\prime+3\vH\vQ+b\,\vH^2\vq+{b\over2}\vH^2_{,\vq}\vq^2+
\vV_{a,\bar q}=0,~~\label{vH1}\\
&\mbox{where}&\vH^2={1\over1-b\vq^2/2}\;\left({1\over2}\vQ^2+\vrho_{{\rm
R}}+\vV_a+\vrho_{{\rm M}}\right). \label{vH} \eea
Here, ``prime'' denotes derivative w.r.t. $\vtau$ and $M$ can be
expressed in terms of the dimensionless quantities a follows
\beq M=\left({(\sqrt{3}\mP)^a\vVo\rho_{\rm
c0}}\right)^{1/(4+a)}~~\mbox{with}~~\vV_a={\vVo/\vq^a}.\label{Mq}\eeq
In our numerical calculation, we use the following values:
\beq \rho_{\rm
c0}\simeq8.1\cdot10^{-47}h^2~\GeV^4,~~\mbox{with}~~h=0.72,~~\vrho_{{\rm
M0}}=0.26 ~~\mbox{and}~~T_0=2.35\cdot 10^{-13}~{\rm GeV}.\eeq
We have also $H_0=2.13\cdot10^{-42}h~{\rm GeV}$ and from
Eq.~(\ref{rhos}) we get $\vrho_{{\rm R0}}=8.04\cdot10^{-5}$.

Eq. (\ref{vH}) can be solved numerically if two initial conditions
are specified at an initial logarithmic time $\vti$, which
corresponds to a temperature $\Ti$ defined as the maximal $T$
after the end of primordial inflation, assuming instantaneous
reheating. We take $q(\vti)=10^{-2}$ throughout our investigation,
without any loss of generality (see below) and let as a free
parameter $\vHi$ (which practically coincides with
$\vQ(\vti)/\sqrt{2}$ since we require a complete domination of
kination at early times as we describe below). To test our model
against observations we extract the density parameters of the
$q$-field, radiation and matter,
\beq \label{omegas}\Omega_i=\rho_i/(\rho_q+\rho_{{\rm
R}}+\rho_{{\rm M}}),~~\mbox{where}~~i=q,~{\rm
R~~\mbox{and}~~M},\eeq
respectively, and the equation-of-state parameter of the
$q$-field, $w_q$,
\beq \label{wq} w_q={P_q\over\rho_q}~~\mbox{where}~~\bar
P_q={1\over2}\vQ^2-\vV_a-{b\over2}\vH^2\vq^2
~~\mbox{and}~~\vrho_q={1\over2}\vQ^2+\vV_a+{b\over2}\vH^2\vq^2,\eeq
with $P_q$ the pressure of $q$ and $\bar P_q=P_q/\rho_{\rm c0}$.

\subsection{Imposed Requirements} \label{reqq}

We impose on our quintessential model the following requirements:

\subsubsection{Constraint of Initial Domination of Kination.}

As stressed in the introduction, we focus on the
range of parameters that ensures the initial domination of the
$q$-kinetic energy. This requirement can be quantified
as follows:
\beq\Omega^{\rm I}_q=\Omega_q(\Ti)=1.\label{domk}\eeq
\subsubsection{Nucleosynthesis Constraint.} The
presence of $\rho_q$ has to respect the successful predictions
of BBN, which commences at about $\vtns=-22.5$
corresponding to $T_{\rm BBN}=1~{\rm MeV}$ \cite{oliven}. Taking
into account the most up-to-date analysis of Ref.~\cite{oliven},
we adopt a rather conservative upper bound on $\Omega_q(\vtns)$,
less restrictive than that of Ref.~\cite{nsb}. Namely, we require:
\beq\Omega_q^{\rm BBN}=\Omega_q(\vtns)\leq0.21~~\mbox{($95\%$
c.l.)} \label{nuc}\eeq
where 0.21 corresponds to additional effective neutrinos species
$\delta N_\nu<1.6$ \cite{oliven}. We do not consider extra
contribution in the left hand side of eq.~(\ref{nuc}), due to the
energy density of the gravitational waves \cite{giova} generated
during a possible former transition from inflation to KD epoch
\cite{qinf}. The reason is that inflation could be driven by
another field different to $q$ and so, any additional constraint
arisen from that period would be highly model dependent.
Nevertheless, inflation can provide a useful constraint for the
parameters of our model as we discuss below.

\subsubsection{Inflationary Constraint.\label{resc}} Resent data
\cite{wmap} strongly favors that the universe underwent an early
inflationary phase. Assuming that this phase is responsible for
the generation of the power spectrum of the curvature scalar
$P_{\rm s}$ and tensor $P_{\rm t}$ perturbations, an upper bound
on the inflationary potential $V_{\rm I}$ and consequently on
$H_{\rm I}$ can be obtained \cite{HIb}. More specifically,
imposing the conservative restriction $r=P_{\rm t}/P_{\rm
s}\lesssim1$, and using the observational normalization of $P_{\rm
s}$ \cite{wmap} we get
\beq H_{\rm I} \lesssim{\pi\over\sqrt{2}}m_{\rm P}P^{1/2}_{\rm s*}
~~\Rightarrow~~H_{\rm
I}\lesssim2.65\cdot10^{14}~\GeV~~\Rightarrow~~\vHi\lesssim1.72\cdot10^{56}
\label{para}\eeq
where $*$ means that  $P_{\rm s*}$ is measured at the pivot scale
$k_*=0.002/{\rm Mpc}$.

\subsubsection{DE-Density and Coincidence Constraint.}
These two requirements can be addressed if (i) the present value
of $\rho_q$, $\rho_{q0}$, is compatible with the preferred range
of \sEref{cdmba}{b} and (ii) $\rho_q$ has already reached the
tracking behavior. The two conditions can be implemented
\cite{mas} if
\beq \mbox{\sf\small
(a)}~~\Omega_{q0}=\vrho_{q0}=0.74~~\mbox{and}~~\mbox{\sf\small
(b)}~~d^2V(\vtau=0)/dq^2\simeq H^2_0,\label{rhoq0}\eeq
where we restrict ourselves to the central experimental value of
$\Omega_{q0}$, since, this choice does not affect crucially our
results on the CDM abundance.

\subsubsection{Acceleration Constraint.} A successful
quintessential scenario has to account for the present-day
acceleration of the universe, i.e. \cite{wmap},
\beq-1.12\leq w_q(0)\leq-0.86~~\mbox{($95\%$ c.l.).}
\label{wqd}\eeq
In our case, we end up with eternal acceleration ($w_q<-1/3$ for
$\vtau>0$).

\paragraph{} Let us finally note that the results obtained on
the age of the universe $t_0$ and the redshift of the transition
from deceleration to acceleration, $z_{\rm t}$,  are marginally
consistent with the experimental data, according to which
$t_0=(13.69\pm0.26)~{\rm Gyr}$ and $z_{\rm
t}=0.46\pm0.26$ at $95\%$ c.l.
We do not impose the experimental data on these quantities as
absolute constraints, due to the observational uncertainties in
their measurement.

\subsection{The Quintessential Dynamics}
\label{Qev}

The cosmological evolution of the various quantities involved in
the model as a function of $\vtau$ is illustrated in \Fref{figw}
for $\vqi=0.01$, $a=0.5$, $b=0.2$, $\Ti=10^9~\GeV$
($\vti\simeq-51.2$) and $\vHi=1.7\cdot10^{52}$ ($\Omqns=0.01$ and
$\vVo=9\cdot 10^{8}$ or $M=4.8~{\rm eV}$). For comparison we also
depict by dotted lines the evolution of certain quantities as a
function of $\vtau$ for $b=0$, $\vVo=8.8\cdot 10^{9}$ (or
$M=8~{\rm eV}$) and the same residual parameters. In particular:

\begin{itemize}

\item In \sFref{figw}{a}, we present $\log\vrho_i$ versus $\vtau$
for  $b=0.2$ (solid lines) and $b=0$ (dotted lines), and for $i=q$
(bold black lines), $i={\rm R+M}$ (light gray lines) and $i={\rm
A}$  (thin black lines). Here, $\vrho_q$ is computed by inserting
in the last equation of Eq.~(\ref{wq}) the numerical solution of
Eq.~(\ref{vH}). The quantity $\vrho_{{\rm R+M}}=\vrho_{{\rm
R}}+\vrho_{{\rm M}}$ is given by Eq.~(\ref{rhos}), and
$\vrho_{{\rm A}}$ is the dimensionless energy density of the
attractor solution (see \Sref{secattr} for details).

\item In \sFref{figw}{b} [\sFref{figw}{c}], we display $q$
[$q^\prime$] versus $\vtau$ for $b=0.2$ (solid line) and $b=0$
(dotted line). We observe that for $b=0$, $q$ grows to a value
greater than $\mP$ before it slows down. As a consequence,
$\vrho_q$ overshoots the tracker solution as shown in
\sFref{figw}{a}. On the contrary, for $b=0.2$ this increase of $q$
can be avoided and the tracker solution is reached before the
present epoch.

\item In \sFref{figw}{d}, we plot $\Omega_i$ -- with $i=q$ (black
line), R (light gray line) and M (gray line) -- and $w_q$ (dark
gray line) versus $\vtau$. We compute $\Omega_i$ [$w_q$] applying
\Eref{omegas} [\Eref{wq}]. We observe that, in the presence of
$q$, the universe undergoes successively a modified KD era, the RD
era and then the matter-dominated era until the re-appearance of
DE. During this KD, $w_q$ takes oscillatory values between $1$ and
$-1$  in sharp contrast to the case of a pure KD era where $w_q=1$
-- see \cref{jcapa}.

\end{itemize}

To further facilitate  the understanding of the quintessential
dynamics we present below a qualitative approach applying the
arguments of \cref{dimopoulos, binetruy}. In particular, $q$
undergoes the following four phases during its evolution:

\begin{figure}[t]\vspace*{-.1in}
\hspace*{-.25in}
\begin{minipage}{8in}
\epsfig{file=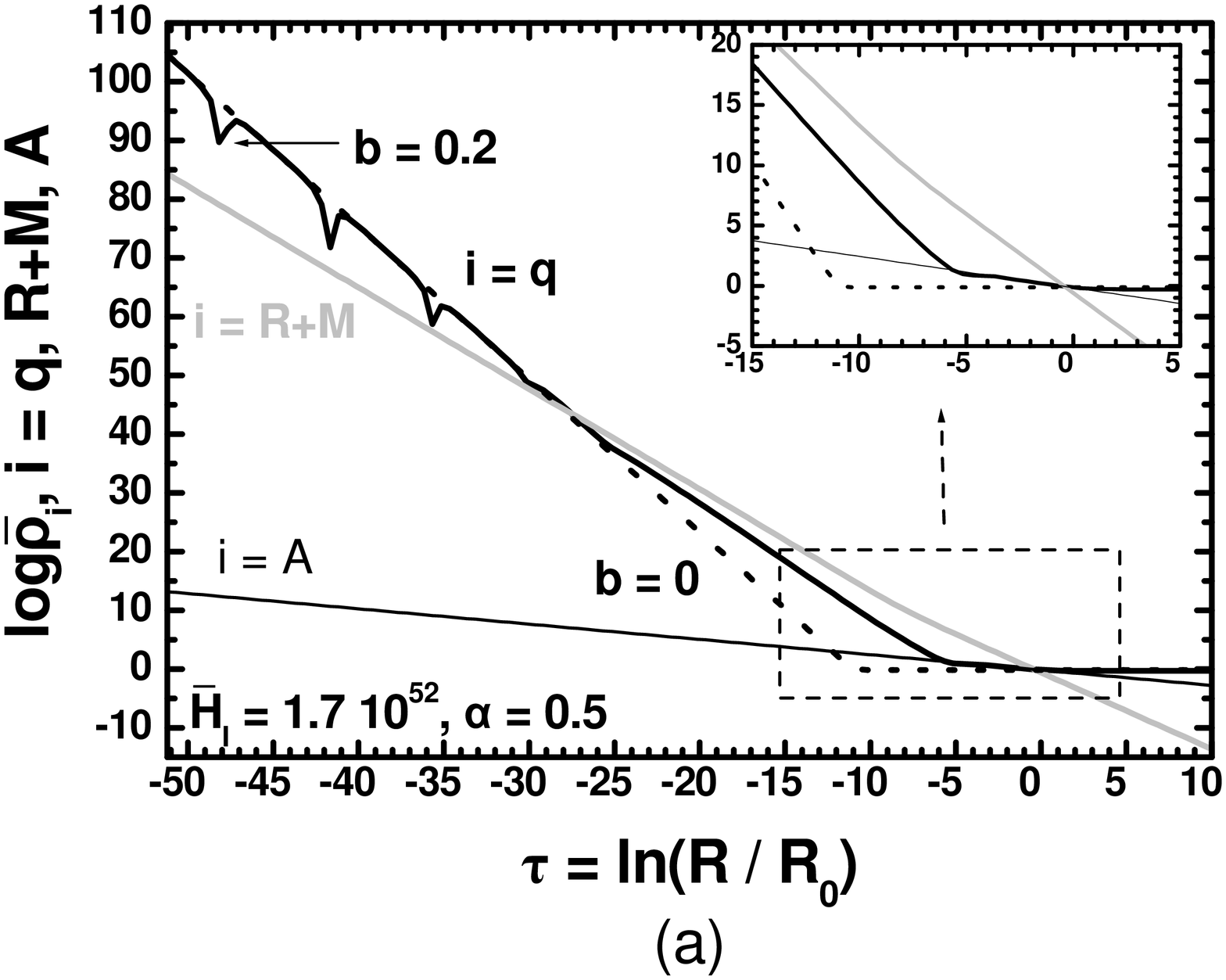,height=3.55in,angle=-90}
\hspace*{-1.37 cm}
\epsfig{file=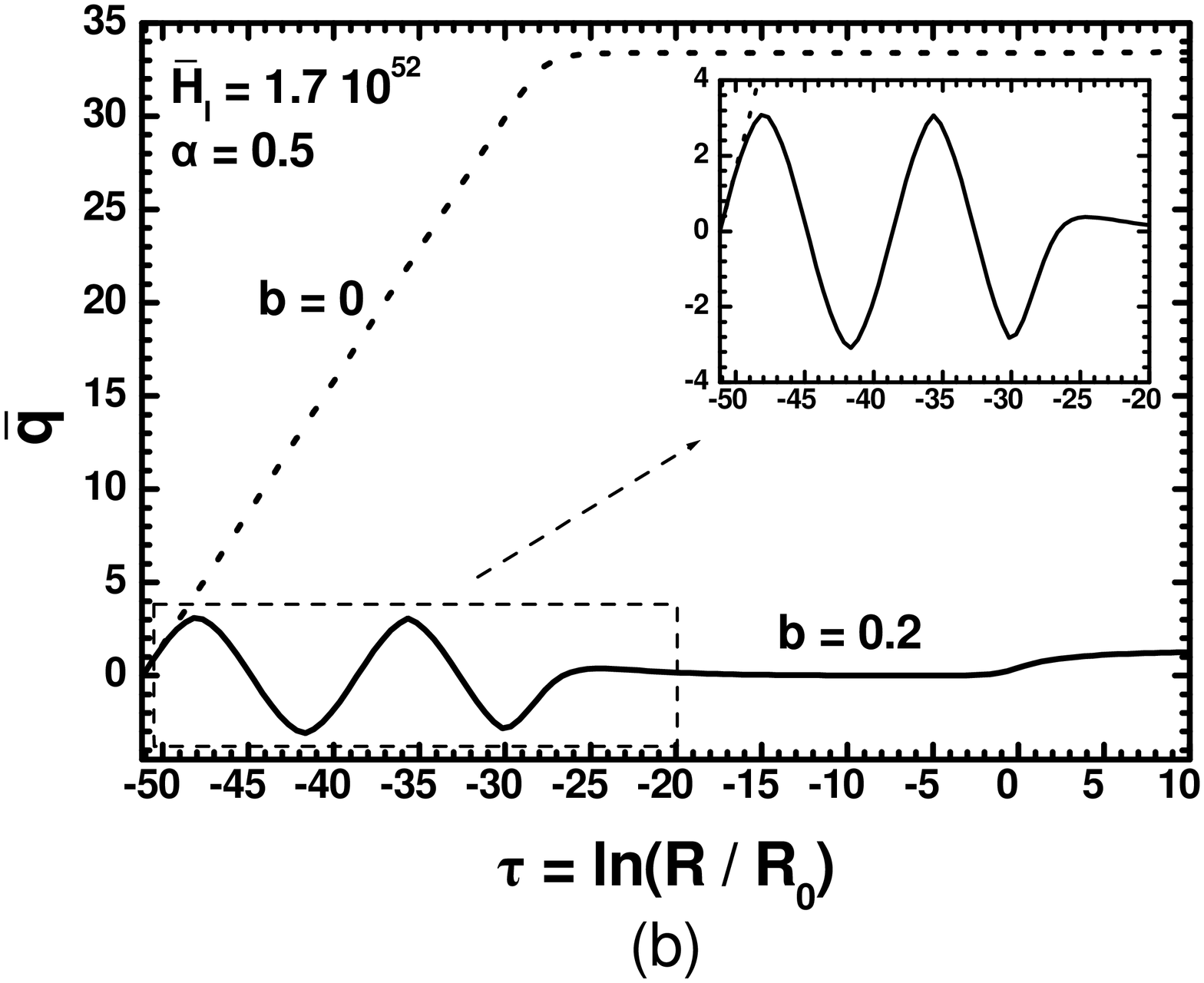,height=3.55in,angle=-90} \hfill
\end{minipage}\vspace*{-.01in}
\hfill \hspace*{-.25in}
\begin{minipage}{8in}
\epsfig{file=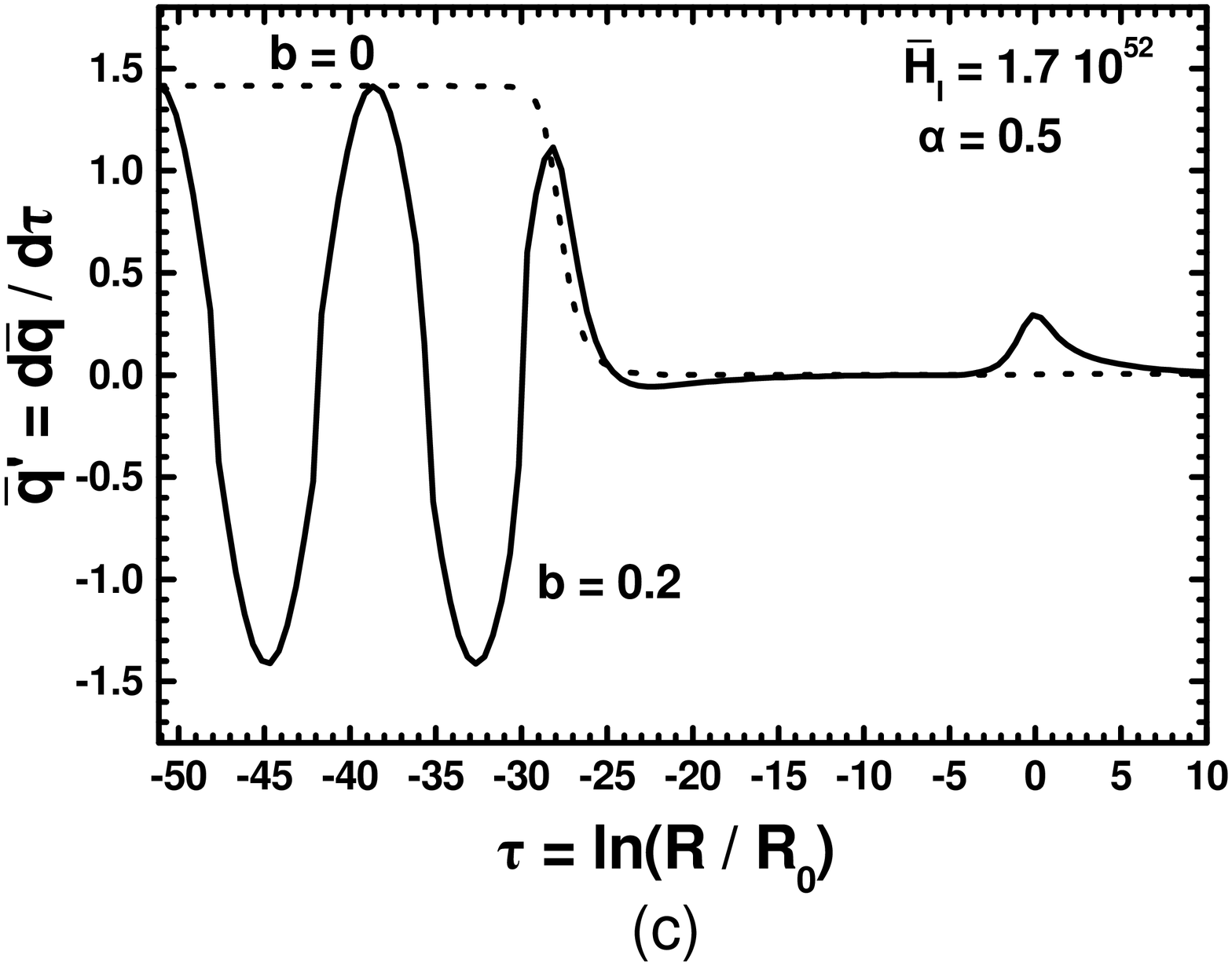,height=3.55in,angle=-90}
\hspace*{-1.37 cm}
\epsfig{file=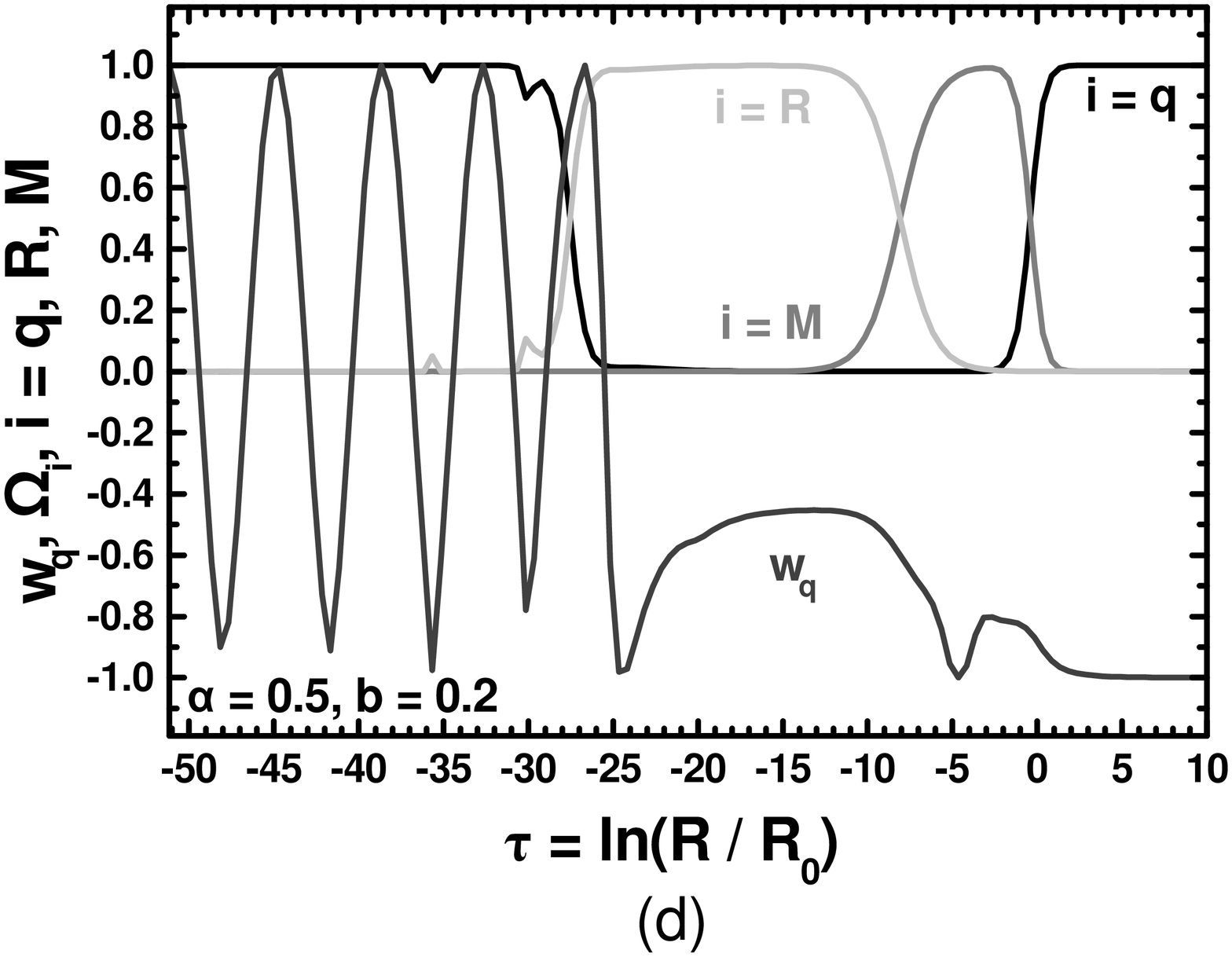,height=3.55in,angle=-90} \hfill
\end{minipage}
\hfill \caption[]{\sl\ftn The cosmological evolution as a function
of $\vtau=\ln(R/R_0)$ for $\vqi=0.01$, $a=0.5$, $b=0.2$,
$\Ti=10^9~\GeV$ ($\vti\simeq-51.2$) and $\vHi=1.7\cdot10^{52}$
($\Omqns=0.01$ and $\vVo=9\cdot 10^{8}$) of the quantities
{\sf\ssz (a)} $\log\vrho_i$ with $i=q$ (solid black line), R+M
(light gray line) and A (thin line) {\sf\ssz (b)} $\vq$ {\sf\ssz
(c)} $\vq'(=d\vq/d\vtau)$ and {\sf\ssz (d)} $w_q$ (dark gray line)
and $\Omega_i$ with $i=q$ (black line), R (light gray line) and M
(gray line). For comparison we also depict by dotted lines the
evolution as a function of $\vtau$ of the quantities {\sf\ssz (a)}
$\log\vrho_q$ {\sf\ssz (b)} $\vq$ and {\sf\ssz (c)} $\vq^\prime$
for $a=0.5$, $b=0$, $\Ti=10^9~\GeV$ ($\vti\simeq-51.2$),
$\vHi=1.7\cdot10^{52}$ and $\vVo=8.8\cdot 10^{9}$.} \label{figw}
\end{figure}

\subsubsection{Kination Dominated Phase.\label{sec:kd}}
During this phase, the evolution of both the universe and $q$ is
dominated by the kinetic-energy density of $q$. Consequently,
\Eref{vH1} reads:
\beq \label{defk}
\vQ^\prime+3\vQ+b\vH\vq\simeq0~~\mbox{and}~~\vQ=\vH\vq'~~\mbox{with}
~~\vH=\sqrt{\vrho_q}\simeq\vQ/\sqrt{2-b\vq^2}.\eeq
The former equations can be integrated trivially to give:
\beq \label{rhok} \vq\simeq\sqrt{2\over
b}\sin\left(\sqrt{b}\,(\vtau-\vti)+\arcsin\sqrt{b\over2}\vqi\right)~~\mbox{and}
~~\vQ\simeq\vQ_{\rm
I}\cos\sqrt{b}(\vtau-\vti)\,e^{-3(\vtauf-\vtif)}. \eeq
Obviously, for $b>0$, $q$ and $Q$ are set in harmonic oscillations
during the KD era. Note that for $b\rightarrow0$, we recover the
well-known results of a pure KD phase \cite{jcapa} -- depicted by
dotted lines in \ssFref{figw}{c}{d}. In other words \cite{mas},
$V$ with $b>0$ acquires a (time-dependent) minimum and $q$ is
prevented from increasing sharply as in the case with $b=0$. In
particular, $\vq$ develops extrema at
\beq \label{tmax}\vtp\simeq\sqrt{1\over
b}\left((2k+1){\pi\over2}-\arcsin\sqrt{b\over2}
\vqi\right)+\vti~~\mbox{with}~~k=0,1,2,...\eeq
On the other hand, $\vq'$ and $\vQ$ almost vanish for $\vtau=\vtp$
(given that $\vqi\sim0$). Therefore, at $\vtau\simeq\vtp$  our
approximation in \Eref{defk} fails instantaneously and $\vrhoR$
dominates over $\vQ^2/2$. As a consequence the $\vq$ and $\vQ$
oscillations become anharmonic and the results of \Eref{rhok}
deviate from their numerical values. These deviations are enhanced
as $|\vtau-\vti|$ increases and/or $\vHi$ decreases. The
oscillatory behavior of $\vq$ is the origin of the peaks shown on
the curve of $\vrho_q$ in \sFref{figw}{a} and the oscillatory form
of $w_q$ in \sFref{figw}{b}. Nevertheless, the simple formula --
see \cref{jcapa} -- which estimates the point $\vtau_{{\rm KR}}$
where the totally KD phase is terminated, is still valid with
rather good accuracy, i.e.,
\beq \label{tkr} \vtau_{{\rm KR}}\simeq\vtau_{_{\rm
I}}+\ln\sqrt{\rho_{q{\rm I}}/{\rho_{{\rm RI}}}}.\eeq
For $b=0.2$ -- and the inputs of \Fref{figw} -- we get
numerically [analytically] $\vtau_{{\rm KR}}=-27.5$ [$\vtau_{{\rm
KR}}=-28$] which corresponds to $T_{{\rm KR}}=0.13~\GeV$ [$T_{{\rm
KR}}=0.2~\GeV$], whereas for $b=0$ the numerical findings
coincide with the analytic ones ($\vtau_{{\rm KR}}=-28$).

\subsubsection{Frozen-Field Dominated Phase.} For $\vtau>\vtkr$, the
universe becomes RD -- and so $\vH^2=\vrho_{{\rm R
}}/(1-b\vq^2/2)$ -- while the evolution of $q$ continues to be
dominated by its kinetic energy density. As a consequence,
\Eref{qeq} is simplified as follows:
\beq \label{deff} \vq''+\vq'+b\,q+{b\over2}\vq\vq'\simeq0. \eeq
Due to the complexity of this equation, it is hard to
to obtain a reliable approximate
solution. What we can say, however, is that, during
this period, both $q$ and $Q$ cease to oscillate and freeze at an
almost constant value. For $b=0.2$, $\log\vrho_q$ decreases
less steeply than for $b=0$ and thus, $\log\vrho_q$ may
join the tracker solution in time -- see \sFref{figw}{a}.

\subsubsection{Attractor Dominated Phase.\label{secattr}}
In this regime, omitting $\ddot q$ and corrections of order $b^2$,
\Eref{qeq} can be simplified as follows:
\beq\label{eomA}\vq'+ {b\over3}\vq -
{a\vVo\over3\vq^{(a+1)}\vrho_{\rm B0}}e^{3(1+w_{\rm
B})\vtau}\simeq0~~\mbox{where}~~\vrhoB=\vrho_{\rm
B0}e^{-3(1+w_{\rm B})\vtau}\eeq
is the dominant background energy density of the universe with
$w_{\rm B}=1/3~[0]$ for the RD~[matter-dominated] era. The
solution of \eref{eomA} can be written as
\beq \label{qA}\vq_{\rm A}\simeq \left({a(a + 2)\vVo\over (9 + (2
+ b) a) \vrhoM_0}\right)^{1/(2 + a)}e^{3(1+w_{\rm B})
\vtau/(a+2)}\eeq
As a consequence \cite{binetruy}, the system in \Eref{qeq} admits
a tracking solution since the energy density of the attractor
\beq \label{rA} \vrho_{\rm A}\simeq\vrho_{\rm Af}e^{-3(1+w_q^{\rm
fp})(\vtau-\vtau_{\rm Af})}~~\mbox{with}~~w_q^{\rm
fp}=\frac{aw_{\rm B}-2}{a+2} \eeq
tracks the dominant $\vrho_{\rm B}$ until $\vtau=\vtau_{\rm Af}$
where the tracking regime finishes and $\vrho_{\rm
B}\simeq\vrho_{\rm A}$. Indeed, $\vrho_A$ decreases less rapidly
w.r.t $\vrho_{\rm B}$ for $a>0$, since
\beq {\vrhoA/\vrhoB}\propto e^{6(1+w_{\rm
B})\vtau/(a+2)}\label{rqrM}\eeq
As a result, $\vrho_q$ eventually dominates and the
expressions leading to the scaling solution of \Eref{rA} can be neglected.
Solving $\vrhoM=\vrhoA$ w.r.t. $\vtau$ (since in our cases
$\vrho_{\rm B}=\vrho_{\rm M}$) we obtain the following expression
for $\vtau_{\rm Af}$:
\beq\vtau_{\rm Af}\simeq \;{1\over6}a\ln{a(2 + a)\over(9 + b\;(2 +
a))} + {1\over3}\ln{\vrhoM_0\over\vVo}\cdot\label{xAf}\eeq

In \sFref{figw}{a} we depict with a thin solid line the evolution
of $\vrho_{\rm A}$ given by \Eref{rA}. For the input parameters of
\Fref{figw} we find that the onset of this phase takes place at
$\vtau_{\rm Ai}\simeq-3.62$ and terminates at $\vtau_{\rm
Af}=-0.4$ with $\vrho_{\rm Af}=0.88$ and $w_q^{\rm
fp}\simeq-0.81$. We check that for $\vtau_{\rm
Ai}\leq\vtau\leq\vtau_{\rm Af}$, $\vrhoA/\vrho_q\simeq0.96$. This
fact -- in conjunction with the fulfillment of \sEref{rhoq0}{b} --
ensures a pure domination of the attractor solution for a
well-defined period, shown in more detail in the subfigure of
\sFref{figw}{a}.

\subsubsection{Vacuum Dominated Phase.} For $\vtau>\vtau_{\rm Af}$, the
universe is dominated by $V$ and so, \Eref{qeq} can be written as
\beq\label{eomP} \vq' + {b\over3}\;\vq - {a\over3\vq}\;
\simeq0~~\Rightarrow~~ \vq\simeq\sqrt{a\over b}\;\left(1 - 9 e^{-2
b \vtau/3} {(a(2 + a))^{ab /9}\over \left(9 + b (2 +
a)\right)^{(1+ab/9)}} \left({\vrhoM_0\over\vVo}\right)^{2
b/9}\right)^{1/2}.\eeq
Using the expression above for $\vq$ we can estimate $w_q(0)$ at
present through the formula
\beq\label{wP} w_q(0)\simeq1-{2\over1+\vq'(0)^{2}/2+b\vq(0)^{2}/2}
\eeq
with results $\vq(0)=0.42$ and $w_q(0)\simeq-0.88$ for the
parameters used in \Fref{figw}. We also obtain $z_{\rm t}=0.76$
and $t_0=13.2~{\rm Gyr}$ which are more or less within the
experimental limits.

\subsection{The Allowed Parameter Space} \label{ap}

\begin{figure}[!t]\vspace*{-.1in}
\hspace*{-.25in}
\begin{minipage}{8in}
\epsfig{file=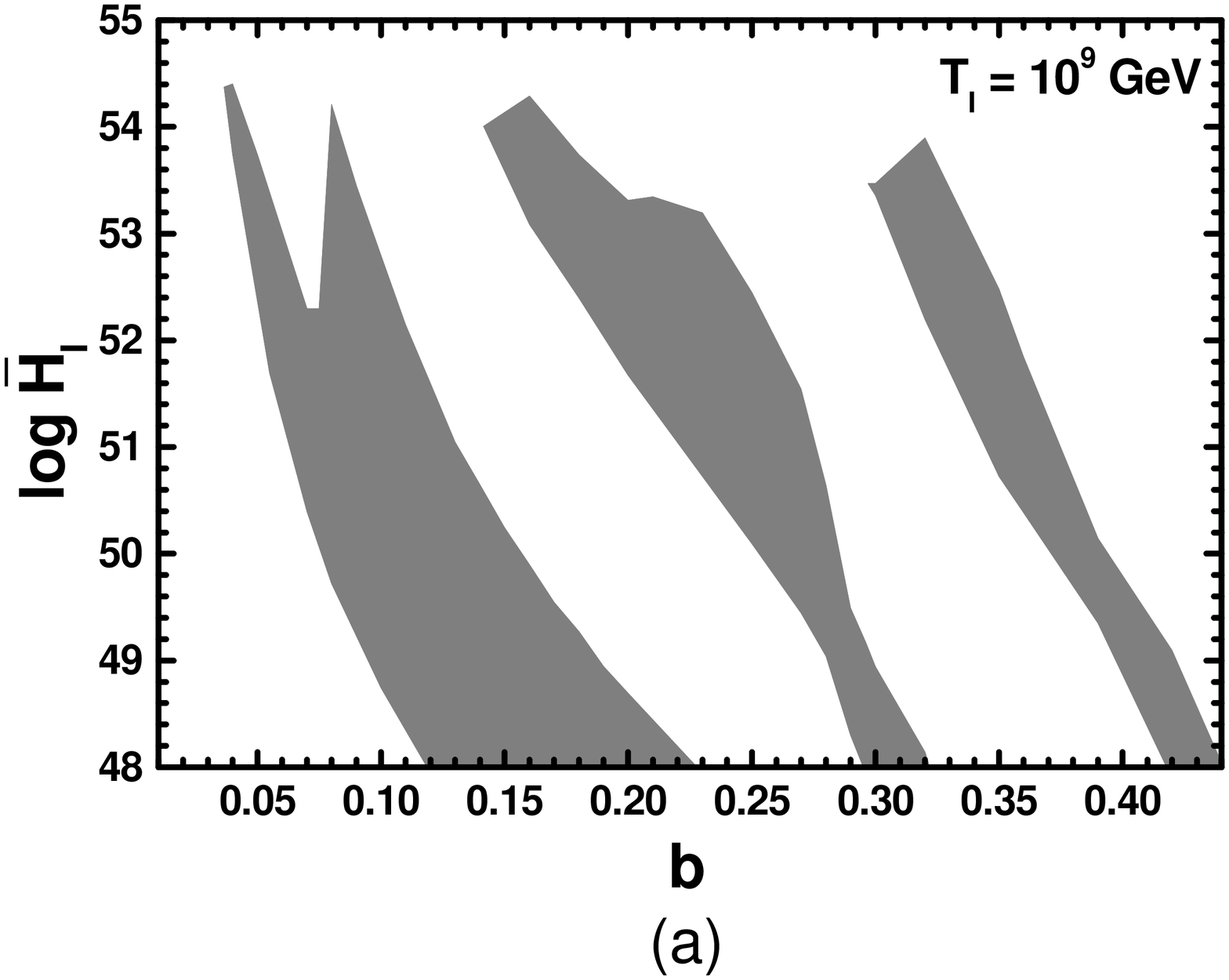,height=3.55in,angle=-90}
\hspace*{-1.37 cm}
\epsfig{file=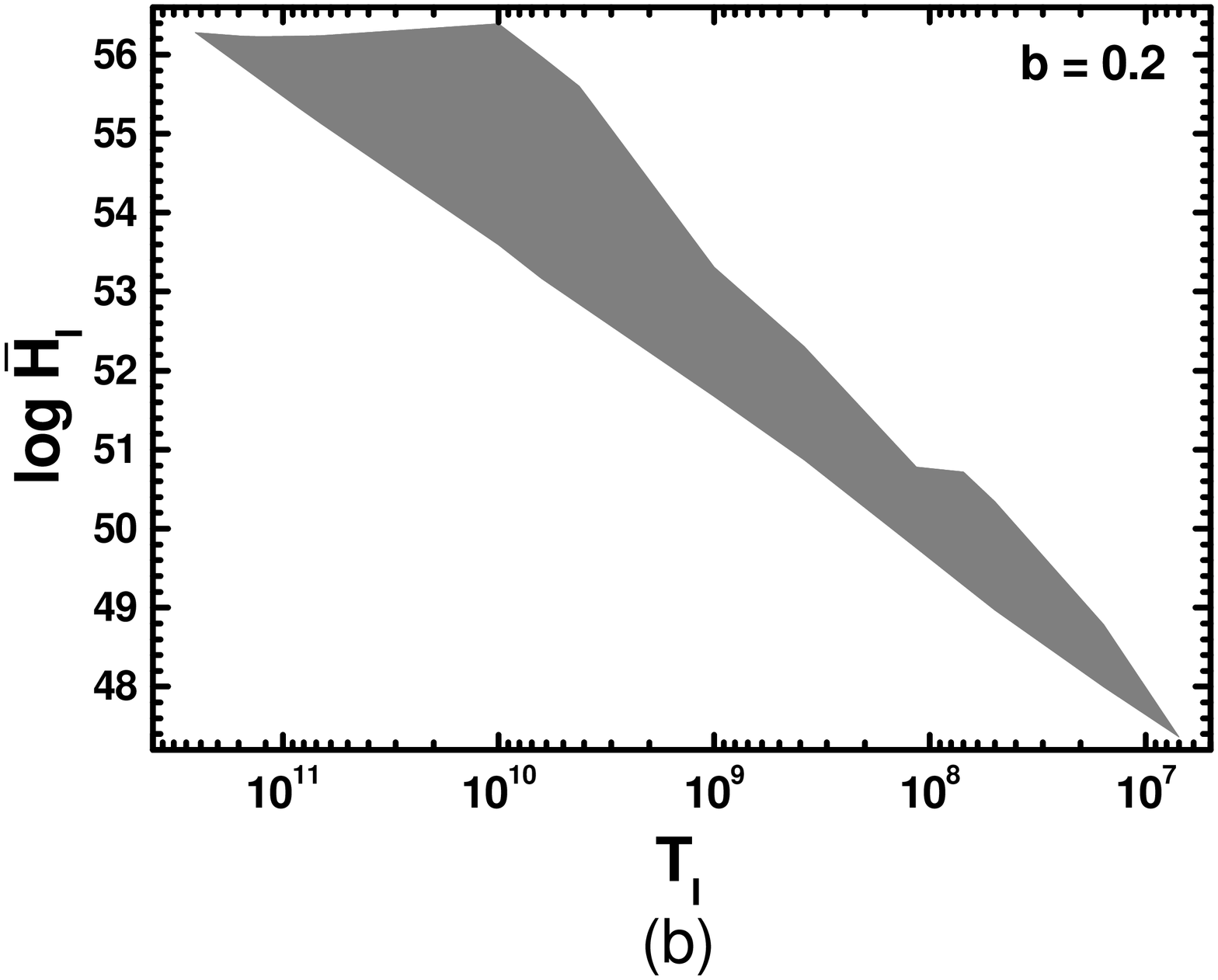,height=3.55in,angle=-90} \hfill
\end{minipage}
\hfill \caption[]{\sl\ftn Allowed (gray shaded) region by
Eqs.~(\ref{domk})-(\ref{wqd}) in the {\sf\ssz (a)} $b-\log\vHi$
plane for $\Ti=10^9~\GeV$ and {\ssz\sf (b)} $\Ti-\log\vHi$ plane
for $b=0.2$. In both cases we take $a=0.5$ and $\vqi=0.01$.}
\label{OmT}
\end{figure}

The free parameters of our quintessential model are:
$$a,\ b,\ M,\ \vti~~(\mbox{or}~~\Ti),\ \vqi~~\mbox{and}~~\vHi.$$
Agreement with Eq.~(\ref{wqd}) implies  $0<a\lesssim0.6$ (compare
also with \cref{bacci}, where a less restrictive upper bound on
$w_q(0)$ has been imposed). The parameter $M$ can be determined
for every $a$ through \Eref{Mq}, so that \sEref{rhoq0}{a} is
satisfied. The determination of $a$ and $M$ is independent of
$\vti,~\vqi$ and $\vHi$, provided that the tracking solution is
reached in time. This property gives an idea of the stability of
the tracking solutions. Note that in the case of the exponential
potential, studied in \cref{jcapa, huelva}, any variation on
$\vqi$ and $\vHi$ requires a re-adjustment of $\vVo$ so that
\sEref{rhoq0}{a} is met. On the other hand, $\Omqns$ (which
influences the calculation of $\Omx$) does depend crucially on
$\vHi$ and $\vti$ (and very weekly on $\vqi$).

In \sFref{OmT}{a} we illustrate the allowed parameter space of our
model in the $b-\log\vHi$ plane for $\Ti=10^9~\GeV$, $a=0.5$ and
$\vqi=0.01$. In the gray shaded areas, Eqs.~(\ref{domk}) -
(\ref{wqd}) are fulfilled. The upper boundary curves of the
allowed bands come from \Eref{nuc}. In the overall allowed region,
we obtain $10^{-6}\lesssim\Omqns\lesssim0.21$. Note, however, that
saturation of \Eref{nuc} is not possible for $0.08<b<0.16$.
Clearly, the model possesses an allowed parameter space with a
band structure. If $(b,\log\vHi)$ belongs in a white [gray] band
the resulting $\vq$ after the oscillatory phase turns out to be
negative [positive] and thus, it cannot [can] serve as
quintessence. Let us once more fix $b=0.2$. For
$51.7\lesssim\log\vHi\lesssim53.3$, $\vq$ develops five extrema
during its evolution -- which is of the type shown in the
fubfigure of \sFref{figw}{b} -- resulting to $\vq_0>0$. As
$\log\vHi$ decreases below $53.3$ (where the bound of \Eref{nuc}
is saturated), the amplitude of the fifth peak, which appears in
the $\vq$-evolution (at about $\vtau\simeq-24.5$) eventually
decreases and finally this peak disappears at $\log\vHi\simeq51.7$
where the first allowed band terminates. For
$48.7\lesssim\log\vHi\lesssim51.7$, $\vq$ develops four extrema
during its evolution, resulting to $\vq_0<0$. As $\log\vHi$
decreases below $51.7$ the amplitude of the forth peak which
appears in the $\vq$-evolution (at about $\vtau\simeq-30$)
decreases and finally this peak disappears at $\log\vHi\simeq48.7$
where the second allowed band commences. Note that in the first
allowed band $\Omqns$ increases with $\vHi$ but this is not a
generic rule (as in the case of a pure KD era).

Fixing $b=0.2$ and letting  $\Ti$ vary in a range of relatively
high values (motivated by the models of SUSY hybrid inflation
\cite{susyhybrid}), we depict in \sFref{OmT}{b} the region allowed
by Eqs.~(\ref{domk}) - (\ref{wqd}) in the $\Ti-\log\vHi$ plane for
$a=0.5$ and $\vqi=0.01$. The upper boundary of the allowed region
comes from \Eref{para} for
$6.7\lesssim\Ti/10^{10}~\GeV\lesssim25.4$ whereas for
$3.8\lesssim\Ti/10^{9}~\GeV\lesssim67$ it arises from the
condition $\vq(0)>0$. The same applies for the left boundary of
the allowed region and several parts of its right boundary. On
some parts of the latter boundary, \Eref{nuc} is also saturated.
In the overall allowed region we obtain
$10^{-9}\lesssim\Omqns\lesssim0.21$.

\section{Thermal Abundance of CDM Candidates}\label{sec:boltz}

We turn now to the calculation of the thermal abundance, $\Omega_X
h^2$ of a CDM candidate, $X$, which can be a WIMP or an
\emph{e}-WIMP. If $X$ is a WIMP we assume that it maintains
kinetic and chemical equilibrium with plasma, is produced through
thermal scatterings and decouples (being non-relativistic) during
the KD epoch. If $X$ is an \emph{e}-WIMP, we expect that its relic
abundance, due to its early decoupling from the thermal bath, is
diluted after inflation at a relatively high energy scale and we
compute its abundance produced through thermal scatterings and
decays during the KD era. In \Sref{Beq} we present the Boltzmann
equation that governs the evolution of the $X$ number density and
then  describe the procedure we employ to solve this equation
numerically (\Sref{Neqs}) and semi-analytically (\Sref{Seqs}).

\subsection{The Boltzmann Equation}
\label{Beq}

Since $X$'s are in kinetic equilibrium with the cosmic fluid,
their number density, $n_{X}$, evolves according to the Boltzmann
equation:
\beq \dot n_X+3Hn_X= \left\{\matrix{
-\sgv\left(n_X^2 - n_X^{\rm eq2}\right)~~\hfill&\mbox{if $X$ is
WIMP}, \hfill \cr
C_X n^{{\rm eq}2}+
\sum_i{g_i\over2\pi^2}m_i^2\,T\,K_1(m_i/T)\,\Gm{i}~~\hfill
&\mbox{if $X$ is \emph{e}-WIMP}, \hfill \cr}
\right. \label{nx}\eeq
where $H$ is given by Eq.~(\ref{rhoqi}). Let us define the
residual symbols of \Eref{nx} separately:

\subsubsection{The Case of WIMPs.} In this case, $\sgv$ is the
thermal-averaged cross section of WIMPs (hereafter denoted as
$\chi$'s) times the velocity and $n_{\chi}^{\rm eq}$ is the
equilibrium number density of $\chi$, which obeys the
Maxwell-Boltzmann statistics:
\begin{equation} \label{neq}
n_{\chi}^{\rm eq}(x)=\frac{g}{(2\pi)^{3/2}}
m_{\chi}^3\>x^{3/2}\>e^{-1/x}P_2\left({1\over
x}\right),~~\mbox{where}~~x={T\over
m_{\chi}}~~\mbox{and}~~P_n(z)=1+{(4n^2-1)\over8z}
\end{equation}
is a function obtained by expanding the modified Bessel function
of the second kind of order $n$ for $x\ll1$ ($\mx$ is the mass of
$\chi$). Assuming that ${\chi}$'s are Majorana fermions, we set
$g=2$ for their number of degrees of freedom. Let us clarify that
$\sgv$ can be derived from $\mx$ and the residual (s)-particle
spectrum, once a specific theory has been adopted. Following our
strategy in \cref{jcapa}, we treat $\mx$ and $\sgv$ as unrelated
input parameters in order to keep our presentation as general as
possible. Note, also that far enough from $s$-poles and thresholds
$\sgv$ can be expanded \cite{jungman} non-relativistically as
$\sgv={\rm a}+{\rm b}x$, where a and b are constants.

\subsubsection{The Case of {\emph e}-WIMPs.} In this case,
$\nequ={\zeta(3)T^3/\pi^2}$ is the equilibrium number density of
the bosonic relativistic species, $m_i$ [$g_i$] is the mass
[number of degrees of freedom] of the particle $i$ and $K_n$ is
the modified Bessel function of the 2nd kind of order $n$. In the
relativistic regime ($T\gg m_i$) $C_\chi$ has been calculated
using the \emph{Hard Thermal Loop Approximation} \cite{Bolz,
steffenaxino}, resulting to $C_X=C_X^{\rm HT}$, where
\beq C_X^{\rm HT} =\left\{\matrix{
\hspace{-.1cm}\begin{minipage}[h]{7cm}\vspace*{-.3cm}
\bea\nonumber {3\pi\over16\zeta(3)m^2_{\rm P}}\sum_{\alpha=1}^{3}
\left(1+{M_\alpha^2\over3 m_{X}^2}\right)c_\alpha g_\alpha^2
        \ln\left({k_{\alpha}\over g_\alpha}\right)~~~~
\eea\end{minipage}~~\mbox{for}  & X=\Gr, \hfill \cr
\hspace{-.1cm}\begin{minipage}[h]{7cm}
\vspace*{-.3cm}\bea\nonumber {108\pi g_a^2 g_3^2\over \zeta(3)}
\ln\left({1.211\over g_3}\right)~~
\mbox{with}~~g_a={g^2\over32\pi^2f_a}\eea\end{minipage} ~~
\mbox{for} & X=\tilde a. \hfill \cr}
\right. \label{sig1}\eeq
Here, $g_\alpha$ and $M_\alpha$ (with $\alpha=1,2,3$) are the
gauge coupling constants and gaugino masses respectively,
associated with the gauge groups $U(1)_{\rm Y}$, $SU(2)_{\rm L}$
and $SU(3)_{\rm C}$, $(k_\alpha)=(1.634,1.312,1.271)$ and
$(c_\alpha)=(33/5,27,72)$. Note that we include the recently
corrected \cite{steffenaxino} nominator ($1.211$) in the logarithm
of $C_{\ax}^{\rm HT}$. Throughout our analysis we impose universal
initial conditions for the gaugino masses, $M_\alpha(M_{\rm
GUT})=M_{1/2}$ and gauge coupling constant unification, i.e.,
$g_\alpha(M_{\rm GUT})=g_{\rm GUT}$ with $M_{\rm
GUT}\simeq2\cdot10^{16}~\GeV$. \Eref{sig1} can be applied
self-consistently only for $g_\alpha(T)<1$ or equivalently
$T>\Tc=10^4~\GeV$. Towards lower $T$'s, non-relativistic ($T\ll
m_i$) contributions and $X$ production from decays start playing
an important role. These contributions have been incorporated in
our computation for the case of $\ax$, following the formalism of
\cref{huelva}. In particular, for $T\ll m_i$,
$C_{\ax}=C_{\ax}^{\rm LT}$ has been calculated numerically and
$\Gm{i}$'s with $i=\gl,~\sq$ and $\tilde B$ are taken into account
\cite{axino, small, huelva} using the following benchmark values
of $m_i$'s:
\beq \label{mi} \left(m_{\sq},~m_{\gl},~m_{\tilde
B}\right)=(1,~1.5,~0.3)~\TeV.\eeq
On the other hand, in the case of $\Gr$, we restrict ourselves to
the high temperature regime (no formalism for the $\Gr$ production
at low $T$ is available to date). We do not include $\Gr$
production from sparticle decays in the plasma \cite{strumia}
which can change \cite{ewimps} $\Omgr$ by a factor of about two
but does not alter our conclusions in any essential way.

\subsection{Numerical Solution}\label{Neqs}

In order to find a precise numerical solution to our problem, we
have to solve \Eref{nx} together with \Eref{qeq}. To this end we
follow the strategy of Sec.~\ref{Beqs}, introducing the
dimensionless quantities:
\bea && \bar n_X=n_X/\rho_{\rm c0}^{3/4},~\bar n^{\rm
eq}_{X}=n^{\rm eq}_{X}/\rho_{\rm c0}^{3/4},~\bar n^{\rm eq}=n^{\rm
eq}/\rho_{\rm c0}^{3/4},~\vsv=\sqrt{3}m_{\rm P}\;\rho_{\rm
c0}^{1/4}\sgv,\\
&& \bar{C}_X=\sqrt{3}m_{\rm P}\;\rho_{\rm c0}^{1/4}C_X,~\bar
m_i={m_i/\rho_{\rm c0}^{1/4}},~\bar T={T /\rho_{\rm
c0}^{1/4}}~~\mbox{and}~~\bar\Gm{i}={\Gm{i}/H_0}.\eea
In terms of these quantities, Eq.~(\ref{nx}) takes the following
master form for numerical manipulations:
\beq \vH\vn^\prime_X+3\vH \vn_{X}=\left\{\matrix{
-\vsv\ \left(\vn_X^2 - \vn_X^{\rm eq2}\right)~~\hfill&\mbox{if $X$
is WIMP}, \hfill \cr
\bar{C}_X\vn^{\rm eq2}+\sum_i{g_i\over2\pi^2}\bar\Gm{i}\bar m_i^2
\bar T\,K_1(m_i/T)~~\hfill &\mbox{if $X$ is \emph{e}-WIMP}, \hfill
\cr}
\right. \label{rx}\eeq
where $\vH$ is given by \Eref{vH}. The integration of \Eref{rx} is
done until $\vtns\simeq-22.5$ (an integration until $0$ gives the
same result). In the case of WIMPs, we impose the initial
condition $\vn_\chi(\vtau_\chi)=\vn_{\chi}^{\rm eq}(\vtau_\chi)$,
where $\vtx$ corresponds to the beginning ($x=1$) of the Boltzmann
suppression of $\vn_{\chi}^{\rm eq}$. In the case of
\emph{e}-WIMPs, we set the initial condition $\vn_X(\vti)\simeq0$.
We use $C_X=C^{\rm HT}_X$ if $\Ti\gg \Tc$ and $\Tkr\gg \Tc$. On
the other hand, if $\Ti\gg \Tc$ and $\Tkr\ll\Tc$ we integrate
successively \Eref{rx} from $\vti$ to $\vts\simeq-37$ -- which
corresponds to $\Ts=1~\TeV$ -- with $C_{\ax}=C^{\rm HT}_{\ax}$ and
then from $\vts$ to $\vtns$ with $C_{\ax}=C^{\rm LT}_{\ax}$.
Finally, $\OmX$ is evaluated from the well-known formula:
\begin{equation}
\label{om1} \Omega_{X}=\rho_{X0}/\rho_{\rm c0}= m_{X} s_0
Y_{X0}/\rho_{\rm c0}~\Rightarrow~\OmX = 2.748 \cdot 10^8\ Y_{X0}\
m_{X}/\mbox{GeV}.
\end{equation}
where $\rho_X=m_X\,n_X$, $Y_X=n_{X}/s$ is the $X$ yield and
$s_0\,h^2 /\rho_{\rm c0}= 2.748 \cdot 10^8/\GeV$.

\subsection{Semi-Analytical Approach}\label{Seqs}

To facilitate the understanding of our results we adapt the
approach carried out in \cref{jcapa, huelva} to our set-up. In
particular, re-expressing \Eref{nx} in terms of the variables
$Y_X$, $Y_X^{\rm eq}=n^{\rm eq}_{X}/s$ and $Y^{\rm eq}=n^{\rm
eq}/s$ (in order to absorb the dilution term) and converting the
derivatives w.r.t $t$ to derivatives w.r.t $\vtau$, we obtain:
\bea \label{BEf}
Y_X^{\prime}=y\sqrt{\frac{g_b}{g_q}}\left\{\matrix{
(-)\vsv\left(Y_X^2 -Y_X^{\rm eq2}\right)~~\hfill&\mbox{if $X$ is
WIMP}, \hfill \cr
\bar{C}_X Y^{\rm eq2}~~\hfill &\mbox{if $X$ is \emph{e}-WIMP},
\hfill \cr}
\right. ~~\mbox{where}~~~~~~~~~~~~~~~~~\\
\nonumber y(\vtau)=\frac{\bar
s}{\sqrt{\vrhoR}}~~\mbox{with}~~\vs={s\over\rho_{\rm
c0}^{3/4}},~~g_b=1-\frac{b}{2}\vq^2~~\mbox{and}~~g_q\simeq\left\{\matrix{
1+\vQ^2/2\vrhoR\hfill &\mbox{for}~~ \vtau\ll\vtkr,\cr
1\hfill & \mbox{for}~~\vtau\gg\vtkr.\hfill \cr}
\right.\eea
In extracting \Eref{BEf} we keep only the first two terms in the
parenthesis of \Eref{vH} and the first term of the left hand side
of \Eref{nx} for \emph{e}-WIMPs (see below). For $g_b=1$ (or
$b=0$) [$g_b=g_q=1$], we reproduce the well-known results of the
pure KD phase [of the SC]. Let us discuss how we can solve
\Eref{BEf} in the two cases separately:

\subsubsection{The Case of WIMPs.} An accurate approximate solution
to \Eref{BEf} can be achieved, introducing the notion of
freeze-out temperature, $T_{\rm F}=T(\vtf)=x_{{\rm F}}m_{\chi}$
(see, e.g., \cref{jcapa} and references therein), which allows us
to study \Eref{BEf} in the two extreme regimes:

\begin{itemize}

\item For $\vtau\ll \vtf$, $Y_\chi\simeq Y_\chi^{\rm eq}$. In this
case, it is more convenient to rewrite eq.~(\ref{BEf}) in terms of
the variable $\Delta(\vtau)=Y_\chi(\vtau)-Y_\chi^{\rm eq}(\vtau)$
as follows:
\beq \label{deltaBE} \Delta^{\prime}=-{Y_\chi^{\rm
eq}}^{\prime}-y\ \vsv\Delta\left(\Delta+2Y_\chi^{\rm
eq}\right)\sqrt{g_b/g_q}. \eeq
The freeze-out logarithmic time $\vtf$ can be defined by
\beq \Delta(\vtf)=\delta_{\rm F}Y_\chi^{\rm eq}(\vtf)~~\Rightarrow
~~\Delta(\vtf)\Big(\Delta(\vtf)+2Y_\chi^{\rm
eq}(\vtf)\Big)=\delta_{\rm F}(\delta_{\rm F}+2)\ Y_\chi^{\rm
eq2}(\vtf), \label{Tf} \eeq
where $\delta_{\rm F}$ is a constant of order one, determined by
comparing the exact numerical solution of \Eref{BEf} with the
approximate one under consideration. Inserting \Eref{Tf} into
\Eref{deltaBE}, we obtain the following equation, which can be
solved w.r.t $\vtf$ iteratively:
\bea \Big(\ln Y_\chi^{\rm eq}\Big)^\prime(\vtf) =-y_{{\rm
F}}\vsv\delta_{\rm F} (\delta_{\rm F}+2) Y_\chi^{\rm
eq}(\vtf)\sqrt{g_b}/\sqrt{g_q}(\delta_{\rm F}+1)~~\mbox{with}
\label{xf}
\\ \label{xfa}y_{{\rm F}}= y(\vtf)
~~\mbox{and}~~\Big(\ln Y_\chi^{\rm
eq}\Big)^{\prime}(\vtau)=x^\prime\left(\frac{1}{x^2}-\frac{3}{2x}-
\frac{g^{\prime}_{s*}}{g_{s*}}+\frac{15}{8P_2(1/x)}\right), \eea
where the $x-\vtau$ dependence can be derived from \Eref{rs}.

\item For $\vtau\gg\vtf$, $Y_\chi\gg Y_\chi^{\rm eq}$ and therefore,
$Y_\chi^2-Y_\chi^{\rm eq2}\simeq Y_\chi^2$. Inserting this into
\Eref{BEf} and integrating the resulting equation from $\vtf$ down
to 0, we arrive at:
\beq \label{BEsol} Y_{\chi 0} = \left(Y_{\chi\rm F}^{-1}+J_{\rm F}
\right)^{-1},~\mbox{where}~~J_{\rm F}= \int_{\vtauf_{{\rm F}}}^{0}
d\vtau\ \sqrt{\frac{g_b}{g_q}}\; y\ \vsv~~\mbox{and}~~Y_{\chi\rm
F} =(\delta_{\rm F} +1)\> Y_\chi^{\rm eq}(\vtf).\eeq
Although not crucial, a choice $\delta_{\rm F}=1.2\mp0.2$ assists
us to better approach the precise numerical solution of
\Eref{BEf}.

\end{itemize}

Inserting \Eref{BEsol} into \Eref{om1} we can obtain $\Omx$
semianalytically. We verify that this result matches well the one
found through the purely numerical integration of \Eref{rx}.

\subsubsection{The Case of \emph{e}-WIMPs.\label{BEewimps}}
\begin{figure}[!t]\vspace*{-.1in}
\hspace*{-.25in}
\begin{minipage}{8in}
\epsfig{file=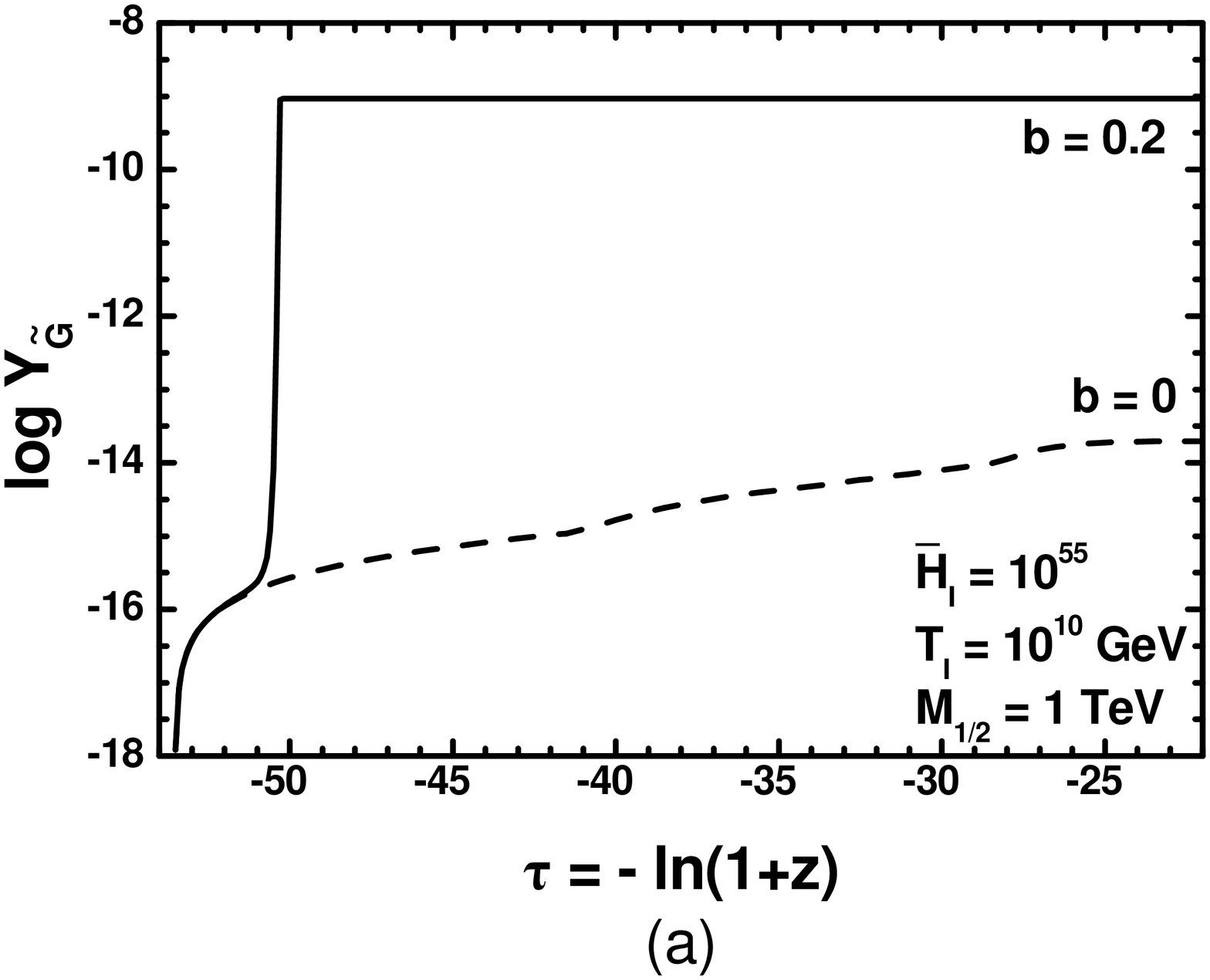,height=3.55in,angle=-90}
\hspace*{-1.37 cm}
\epsfig{file=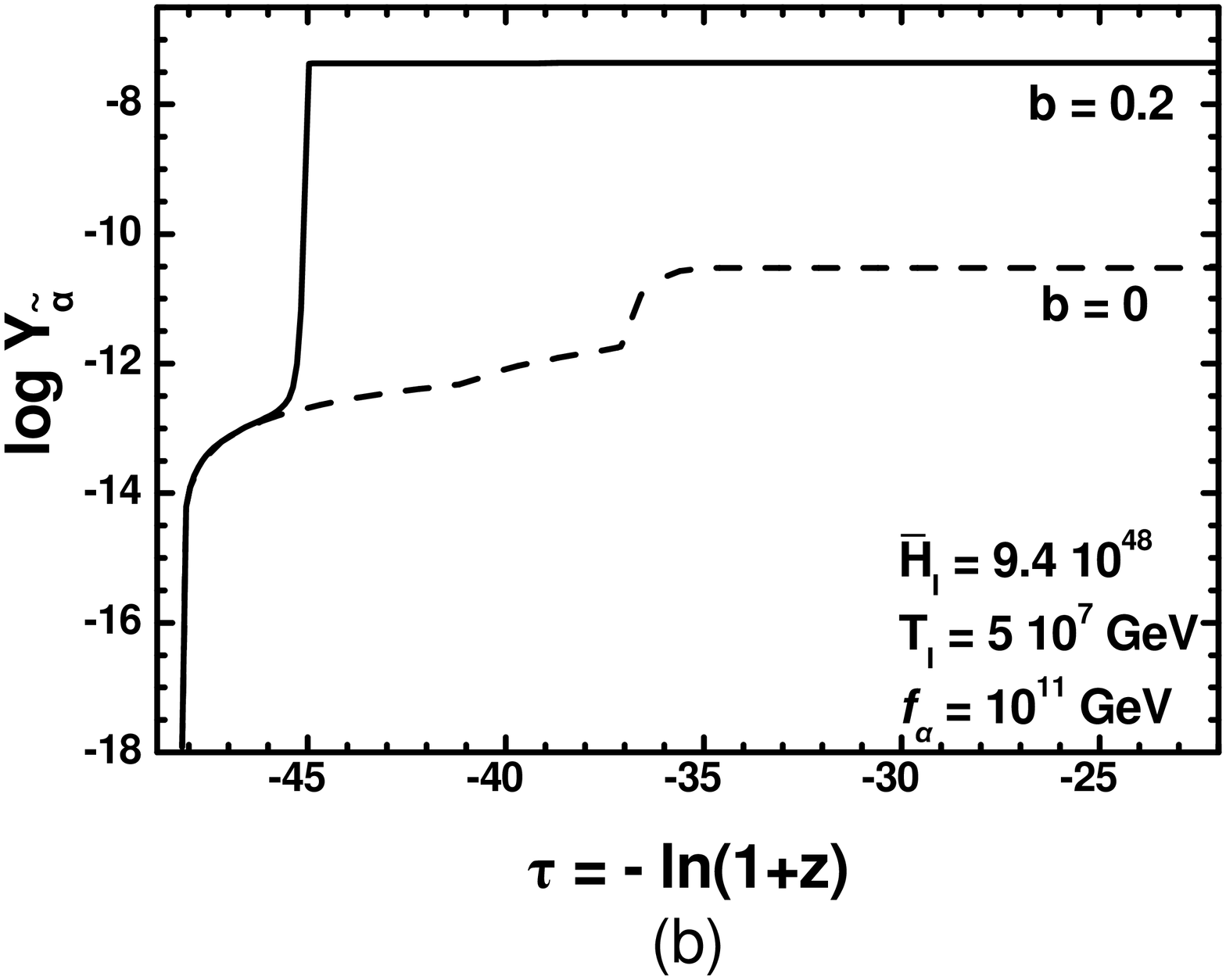,height=3.55in,angle=-90} \hfill
\end{minipage}
\hfill \caption[]{\sl\ftn The evolution of the logarithm of the
$X$ yield, $Y_X$ with $X=\Gr$ [$X=\ax$] as a function of $\vtau$
{\ssz\sf (a)} [{\ssz\sf (b)}] for $a=0.5,~\Ti=10^{10}~\GeV$
[$\Ti=5\cdot10^{7}~\GeV$], $\vHi=10^{55}$
[$\vHi=9.4\cdot10^{48}$], $M_{1/2}=1~\TeV$ [$f_a=10^{11}~\GeV$]
and $b=0$ (dashed line) or $b=0.2$ (solid line). We obtain
$\OmX=0.11$ for $b=0.2$ and $\mgr=0.44~{\GeV}$ [$\mxx=9.5~{\rm
MeV}$].}\label{Ys}
\end{figure}

In this case, we focus on the most intriguing possibility, in
which $\Ti\gg\Ts$ but $\Tkr\ll\Ts$. Under this assumption, $\Ti$
takes sufficient high values, as suggested by the majority of the
inflationary modes (see, e.g., \cref{susyhybrid}) and $\Omqns$
naturally takes values close to its upper bound in \Eref{nuc}.
$Y_{X0}$ can be derived by numerically integrating \Eref{BEf} from
$\vtau=\vti$ until $\vtns$ with $C_X=C^{\rm HT}_X$. This is,
because $Y_{X}$ is stabilized to its final value, $Y_{X0}$, not at
the onset of the RD era -- as in the case of a pure KD era
\cite{huelva} -- but at a high temperature corresponding to $\vtp$
for $k=0$ in \Eref{tmax}. There, the first peak (for $\vtau>\vti$)
of the $q$ evolution takes place and therefore, $\vrhoR$ dominates
over $\vrho_q$ instantaneously as explained in \Sref{sec:kd}.
Consequently, $Y_{X0}$ decreases in our QS w.r.t its value in the
SC but increases w.r.t its value in the case of a pure KD phase.
To illustrate this key point we display in \sFref{Ys}{a}
[\sFref{Ys}{b}] the evolution of $\log Y_X$ for $X=\Gr$ [$X=\ax$]
as a function of $\vtau$ for $\Ti=10^{10}~\GeV$
[$\Ti=5\cdot10^{7}~\GeV$], $\vHi=10^{55}$
[$\vHi=9.4\cdot10^{48}$], $M_{1/2}=1~\TeV$ [$f_a=10^{11}~\GeV$]
and $b=0$ (dashed line) or $b=0.2$ (solid line). For $b=0.2$ and
$\mgr=0.44~{\GeV}$ [$\mxx=9.5~{\rm MeV}$] we get $\OmX=0.11$ (with
$X=\Gr$ [$X=\ax$]).

In \sFref{Ys}{a} [\sFref{Ys}{b}] we observe that for $b=0$, where
a pure KD era occurs, $Y_{X}$ with $X=\Gr$ [$X=\ax$] takes its
final value, more or less, close to the onset of the RD era for
$\vtkr\simeq-28.5$ [$\vtkr\simeq-25.4$] -- corresponding to
$\Tkr\simeq0.23$ [$\Tkr\simeq0.017$] --, according to the
well-known results of \cref{huelva}. Namely, for $X=\Gr$ the
resulting $Y_{X0}$ (for $b=0$) is not so precise, since the used
$C^{\rm HT}_X$ from \Eref{sig1} gives reliable results mainly for
$T\gg\Tc$. On the other hand, for $X=\ax$ (and $b=0$), $Y_{X}$ is
calculated employing for $T<\Ts$, $C_{\ax}=C^{\rm LT}_{\ax}$
obtained from the non-relativistic formalism of \cref{huelva} with
the $m_i$'s indicated in \Eref{mi}. Due to the Boltzmann
suppression occurred for $T<\Ts$, $Y_{X}$ takes its present value
at about $T\simeq\Ts$ ($\vts=-37$). On the contrary, for $b=0.2$
and the inputs of \sFref{Ys}{a} [\sFref{Ys}{b}], $Y_{X}$ takes its
final value at $\vtau\simeq-50$ [$\vtau\simeq-45$] where $q$
develops its first -- for $k=0$ in \Eref{tmax} -- extremum.
Obviously, $Y_{X0}$ for $b=0.2$ is much larger than its value for
$b=0$ but still lower than its value in the SC for $T\simeq\Ti$.
Indeed, for the inputs of \sFref{Ys}{a} [\sFref{Ys}{b}] we obtain
$\log Y_{\Gr 0}=-5.3$ [$\log Y_{\ax 0}=-3.94$] in the SC.

Despite the fact that $C^{\rm HT}_X$ in \Eref{BEf} has a rather
simple form given by \Eref{sig1}, it is not straightforward to
find a general analytical result for the integration of
\Eref{BEf}. This is mainly due to the fact that the integrand
includes $g_q$ and $g_b$ which depend on the $q$ and $Q$ evolution
in a rather complicate way. Nevertheless, we can write simple
empirical relations which reproduce rather accurately (within
$5\%$) our numerical results for $X=\Gr$ [$X=\ax$], $a=0.5$,
$b=0.2$, $\Ti=10^{10}~\GeV$ or $\Ti=10^{9}~\GeV$
[$\Ti=10^{9}~\GeV$ or $\Ti=5\cdot10^{7}~\GeV$]. Namely
\beq\label{emp}\OmX=\Bigg[\left(a_{1X}\log{\Ti\over\GeV}+b_{1X}\right)\log\vHi+
\left(a_{2X}\log{\Ti\over\GeV}+b_{2X}\right)\Bigg]\left\{\matrix{
M^2_{1}/\mP m_X\hspace*{-.2cm}&\mbox{for $X=\Gr$}, \hfill \cr
\mP m_X/f_a^2&\mbox{for $X=\ax$}, \hfill \cr}
\right. \eeq
where the numerical coefficients $a_{1X}$, $a_{2X}$, $b_{1X}$,
$b_{2X}$ are given by
\bea\nonumber\left(a_{1X}, a_{2X}, b_{1X},
b_{2X}\right)=\left\{\matrix{
(-1.603,~ 95.26,~
1.4223,~-845.78)\cdot10^{11}\hspace*{-.2cm}&\mbox{for $X=\Gr$},
\hfill \cr
(-1.2505, ~71.064,~ 9.5244,~ -541.6)\cdot10^{5}&\mbox{for
$X=\ax$.} \hfill \cr}
\right. \eea

We, thus, conclude that in our QS, for  a naturally
expected hierarchy among $\Ti$, $\Tc$ and $\Tkr$, the calculation of
$\OmX$ depends exclusively on $C^{\rm HT}_{X}$ and not on $C^{\rm
LT}_{X}$, $\Gamma_i$'s and $m_i$'s.

\section{CDM from Thermal Production of WIMPs}\label{sec:wimp}

Employing the formalism developed in the previous section, we can
analyze the behavior of $\Omx$ as a function of the free
parameters of the QS (\Sref{subsec1:wimp}) and construct regions
allowed by the various constraints (\Sref{subsec2:wimp}).

\subsection{The Enhancement of $\Omx$}\label{subsec1:wimp}

\begin{figure}[!t]\vspace*{-.1in}
\hspace*{-.25in}
\begin{minipage}{8in}
\epsfig{file=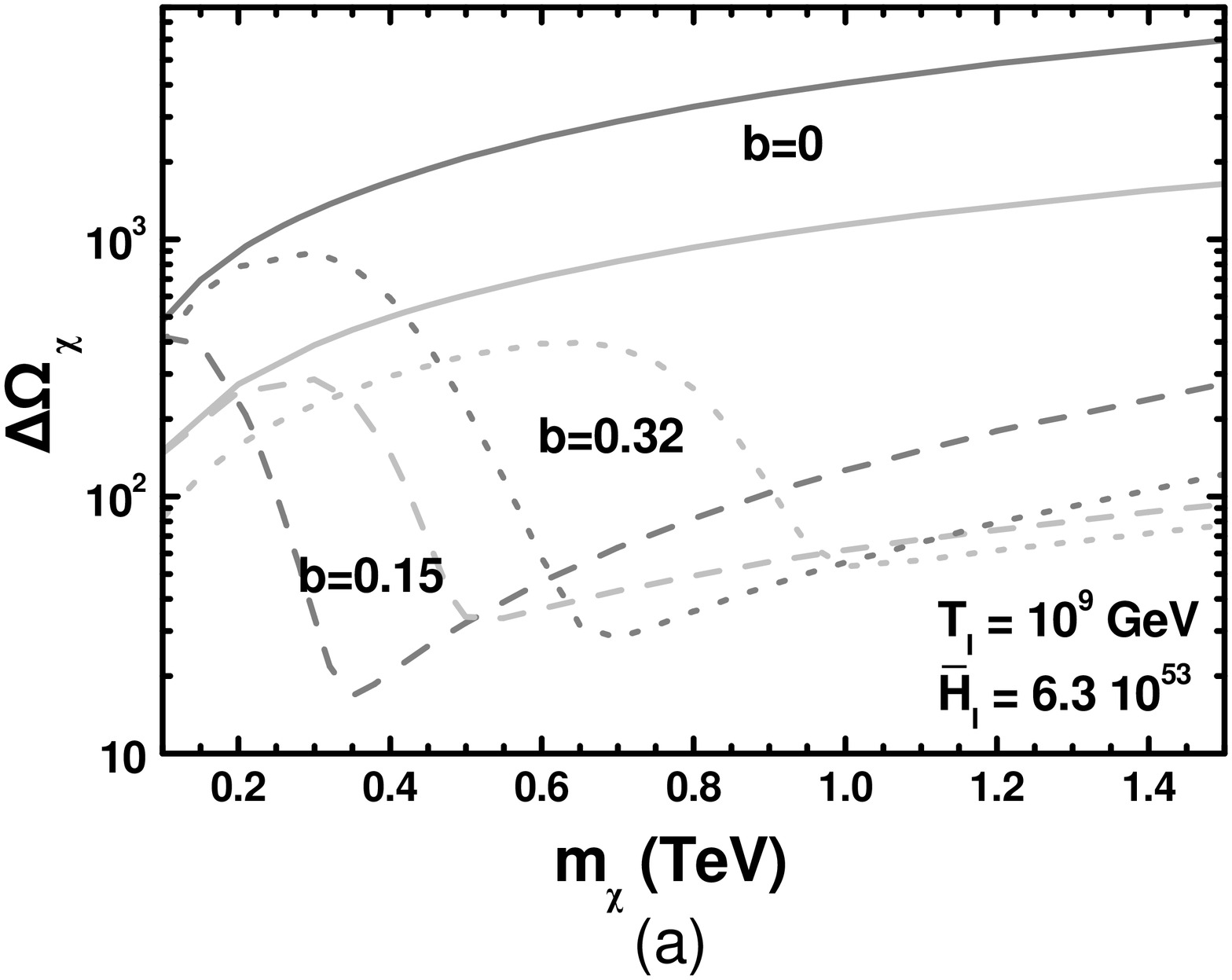,height=3.55in,angle=-90}
\hspace*{-1.37 cm}
\epsfig{file=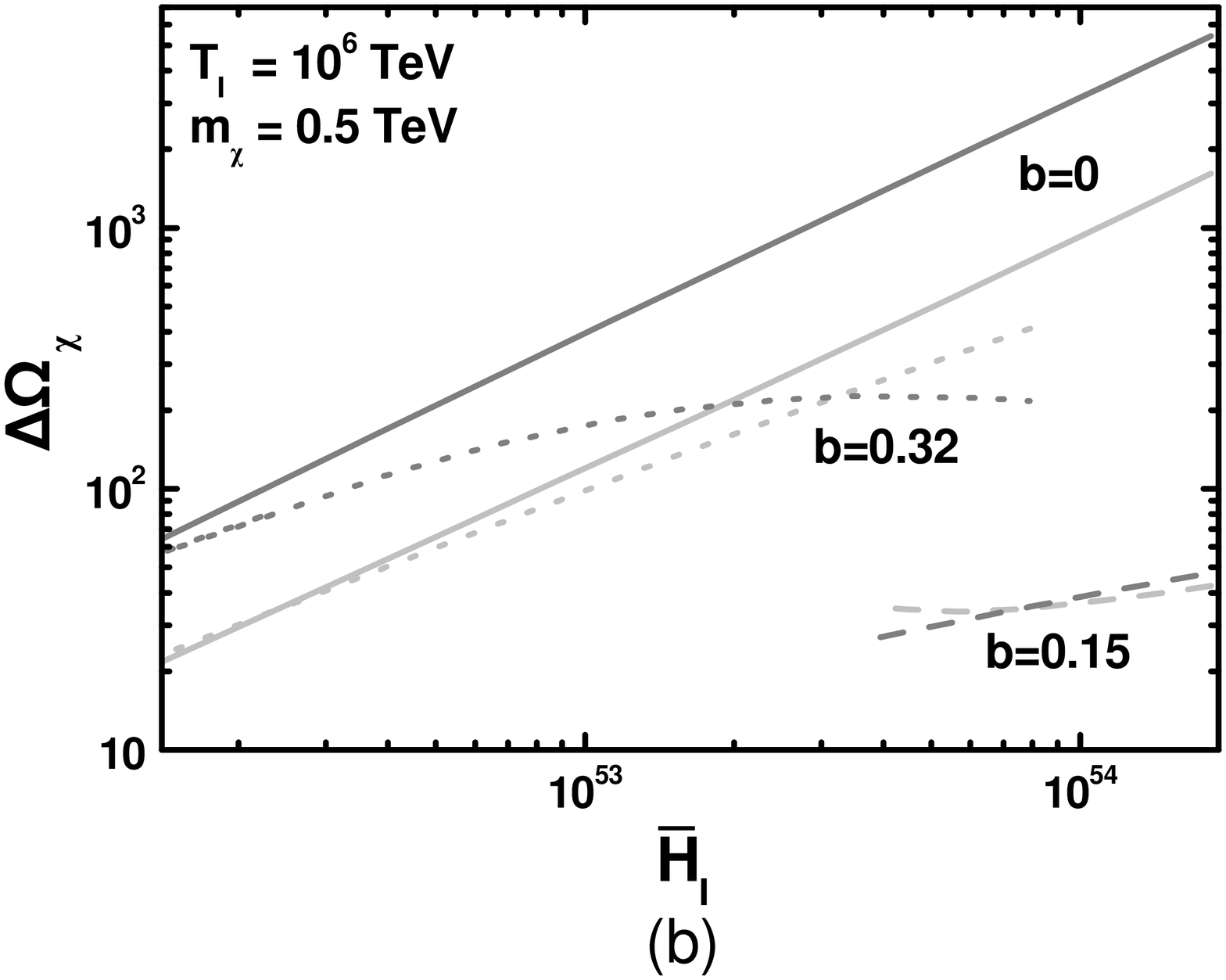,height=3.55in,angle=-90} \hfill
\end{minipage}
\hfill \caption[]{\sl\ftn  $\Delta\Omega_{\chi}$ versus {\sf\ssz
(a)} $m_\chi$ for $\vHi=1.6\cdot10^{53}$ and {\sf\ssz (b)} $\vHi$
for $m_\chi=0.5~\TeV$. In both cases we take
$a=0.5,~\vTi=10^9~\GeV$ (or $\vti=-51.16$), $\langle\sigma
v\rangle=10^{-8}~\GeV^{-2}~[\langle\sigma v\rangle=10^{-10}x~{\rm
GeV}^{-2}]$ (light gray [gray] lines) and $b=0$ (solid lines),
$b=0.15$ (dashed lines) and $b=0.32$ (dotted lines).}\label{om}
\end{figure}

The presence of $g_q>1$ in \Eref{xf} and, mainly, in \Eref{BEsol}
reduces $J_{\rm F}$, thereby increasing $\Omx$ w.r.t its value,
$\left.\Omx\right|_{{\rm SC}}$, in the SC -- i.e., setting
$g_q=g_b=1$ in \Eref{BEf}. The resulting enhancement can be
estimated, by defining the quantity:
\beq\label{dom}\Domx= {\Omx-\left.\Omx\right|_{\rm
SC}\over\left.\Omx\right|_{\rm SC}}\cdot\eeq

The behavior as a function of our free parameters of
$\Delta\Omega_{\chi}$ within our QS can be inferred from \Fref{om}
where we display $\Domx$ versus $\mx$ -- see \sFref{om}{a} -- for
$\vHi=6.3\cdot10^{53}$ or $\vHi$ -- see \sFref{om}{b} -- for
$\mx=0.5~\TeV$. We isolate two cases commonly encountered in the
analysis of several models, fixing $\sgv=10^{-8}~\GeV^{-2}$ (light
gray curves) and $\sgv=10^{-10}x~\GeV^{-2}$ (gray curves). We take
$b=0$ (solid lines), $b=0.15$ (dashed lines) and $b=0.32$ (dotted
lines). The chosen $\vHi$ results to $\Omqns\simeq0.01,~0.068$ or
$0.19$ for $b=0, 0.15$ or $0.32$ correspondingly, whereas in
\sFref{om}{b} $\Omqns$ ranges from $5\cdot10^{-6}$ to $0.079$ for
$b=0$ or from $0.05$ to $0.21$ [from $0.016$ to $0.21$] for
$b=0.15$ [$b=0.32$]. Due to the band structure of the allowed
parameter space of the model -- see \sFref{OmT}{a} -- not all
$\Omqns$'s are achievable for any $b$.

Clearly, for $b=0$ we get a pure KD era and our results reduce to
those presented in \cref{jcapa}, i.e., $\Domx$ increases when
$\mx$ or $\vHi$ (and consequently $\Omqns$) increases or when
$\sgv$ decreases -- see \ssFref{om}{a}{b}. On the contrary, for
$b\neq0$, $\Domx$ depends crucially on the hierarchy between
$\vtf$ and $\vtp$. Given that $J_{\rm F}$ takes its main
contribution from $g_q$ for $\vtau\sim\vtf$, $J_{\rm F}$ is
enhanced -- see \Eref{BEsol} -- if $\vtf$ is lower than $\vtp$ and
close to it, since $g_q$ is suppressed ($g_q\simeq1$) for
$\vtau\simeq\vtp$. As a consequence -- see \eqs{om1}{BEsol} --
$\Domx$ diminishes. This argument is highlighted by
\Tref{Dommin}. There, we list the range of $\vtf$ for
$0.1\leq\mx/\TeV\leq1.5$ and $\sgv=10^{-8}~\GeV^{-2}$ or
$\sgv=10^{-10}x~\GeV^{-2}$ and the logarithmic time $\vtp$ at
which the closest to $\vtf$'s peak in the $q$-evolution takes
place for $b=0.15$ or $b=0.32$ and $\vHi=6.3\cdot10^{53}$.
Clearly, $\vtf$ [$\vtp$] is independent of $b$ or $\vHi$ [$\mx$
or $\sgv$]. As $\vtf$ moves closer to $\vtp$, $\Domx$ decreases
with its minimum $\left.\Domx\right|_{\rm min}$ occurring at
$\vtf\simeq \vtau_{\rm F}^{\rm min}$ or $\mx=m_{\chi}^{\rm min}$.
Note that $\vtau_{\rm F}^{\rm min}$ does not coincide with $\vtp$
always due to the presence of $Y_{\rm F}$ in \Eref{BEsol}. The
appearance of minima can be avoided if $\vtf$'s happen to remain
constantly lower than $\vtp$'s.

\renewcommand{\arraystretch}{1.2}
\begin{table}[!t]
\begin{center}\begin{tabular}{|c|c||c|c|c||c|c|c|} \hline
$\sgv$&$-\vtf$&\multicolumn{3}{|c||}{$b=0.15$,
$\vtp\simeq-33.1$}&\multicolumn{3}{|c|}{$b=0.32$,
$\vtp\simeq-33.8$}\\\cline{3-8}
$\left(\GeV^{-2}\right)$&&$\left.\Domx\right|_{\rm
min}$&$\vtau_{\rm F}^{\rm min}$&$m_\chi^{\rm
min}/\TeV$&$\left.\Domx\right|_{\rm min}$&$\vtau_{\rm F}^{\rm
min}$&$m_\chi^{\rm min}/\TeV$\\\hline \hline
$10^{-8}$ &$ 31.8-34.6$&$32.6$&$-33.4$&$0.52$&$53.2$&$-34.15$&$1.01$\\
$10^{-10}x$& $32.3-35.2$
&$16.8$&$-33.5$&$0.35$&$28$&$-34.3$&$0.69$\\\hline
\end{tabular}
\end{center}\vspace*{-.155in}
\caption{\sl\ftn The minima of $\Domx,~\left.\Domx\right|_{\rm
min}$, occurring at the freeze-out logarithmic time $\vtau_{\rm
F}^{\rm min}$ or the mass of $\chi$ $m_\chi^{\rm min}$,  for
$a=0.5$, $\vTi=10^9~\GeV$, $\vHi=6.3\cdot10^{53}$ and several
$\sgv$'s and $b$'s employed in \sFref{om}{\ssz a}. We also show
the range of $\vtf$ for $0.1\leq\mx/\TeV\leq1.5$ and the
logarithmic time $\vtp$ at which the closest to $\vtf$ peak in the
$q$-evolution takes place.}\label{Dommin}
\end{table}
\renewcommand{\arraystretch}{1.0}

Increasing $\vHi$ for fixed $\mx=0.5~\TeV$ leads to an increase of
$\Omqns$ for the parameters of \sFref{OmT}{b}. However, the
expected increase of $\Domx$ is less effective for
$\sgv=10^{-10}x~\GeV^{-2}$ than for $\sgv=10^{-8}~\GeV^{-2}$ as
shown in \sFref{OmT}{b}. Let us check, e.g., the case for
$b=0.32$. As $\vHi$ increases in the first allowed band, shown in
\sFref{OmT}{a}, from $52.2$ to $53.9$, $\vtp$ moves form $-34$ to
$-33.8$ and influences $\Domx$ for $\sgv=10^{-10}x~\GeV^{-2}$ more
than for $\sgv=10^{-8}~\GeV^{-2}$. This is, because for
$\sgv=10^{-10}x~\GeV^{-2}$, we get $\vtf\simeq-33.7$ which is
closer to $\vtp$'s than $\vtf\simeq-33.1$ which is taken for
$\sgv=10^{-8}~\GeV^{-2}$. Variation of $\Ti$ (or equivalently
$\vti$) leads to a displacement of $\vtp$'s -- see \Eref{tmax} --
and therefore the minima of $\Domx$ in \sFref{OmT}{a} or the
limits of the non solid lines in \sFref{OmT}{b} are relocated.
However, our results on the behavior of $\Domx$ remain intact.

\subsection{Restrictions in the $\mx-\sgv$ Plane}\label{subsec2:wimp}

Though the post-freeze-out $Y_\chi$ in \Eref{BEsol} stays
essentially unchanged, residual annihilations of $\chi$'s occur up
to the present, with several cosmological consequences. Recently,
tight upper bounds on $\sgv$'s have been reported. These, however,
depend on the identity of the products of the annihilation of
$\chi$'s. To get a feeling of the relevant effects, we adopt the
most restrictive bound which arises for the annihilation mode of
$\chi$'s to $\ps\el$. In particular, the constraints from BBN
\cite{moroiBBN} and cosmic microwave background \cite{CMB} result
to the following bounds, respectively:
\beq\label{sigBBN} \mbox{\sf\small (a)}~~
\sgv\leq3\cdot10^{-5}~\GeV^{-2}\;{m_{\nt}\over1~\TeV}
~~\mbox{and}~~\mbox{\sf\small
(b)}~~\sgv\leq4.4\cdot10^{-7}~\GeV^{-2}\;{m_{\nt}\over1~\TeV}\cdot\eeq
Obviously the constraint of \sEref{sigBBN}{b} is much more
restrictive than this from \sEref{sigBBN}{a}. To be in harmony
with the assumptions considered in the derivation of the above
bounds, we consider hereafter $\sgv$'s independent of $T$.

Having fixed the parameters which determine the QS, we can derive
restrictions on the parameters involved in the $\Omx$ calculation.
This is done in \Fref{svmx} where we show the allowed
parameter space in the $m_\chi-\sgv$ plane for $a=0.5$ and
$\vTi=10^9~\GeV$. We also take $b=0$ and $\vHi=6.3\cdot 10^{53}$
or $6.2\cdot10^{52}$ resulting to $\Omqns=0.01$ or $10^{-4}$
respectively in \sFref{svmx}{a}, $\vHi=6.3\cdot10^{53}$ and
$b=0.15$ or $\vHi=6.2\cdot10^{52}$ and $b=0.32$ yielding
$\Omqns=0.068$ or $0.065$ respectively, in \sFref{svmx}{b}.

\begin{center}  \vspace*{-1.1cm}\hspace*{-.25in}
\begin{minipage}{8in}
\epsfig{file=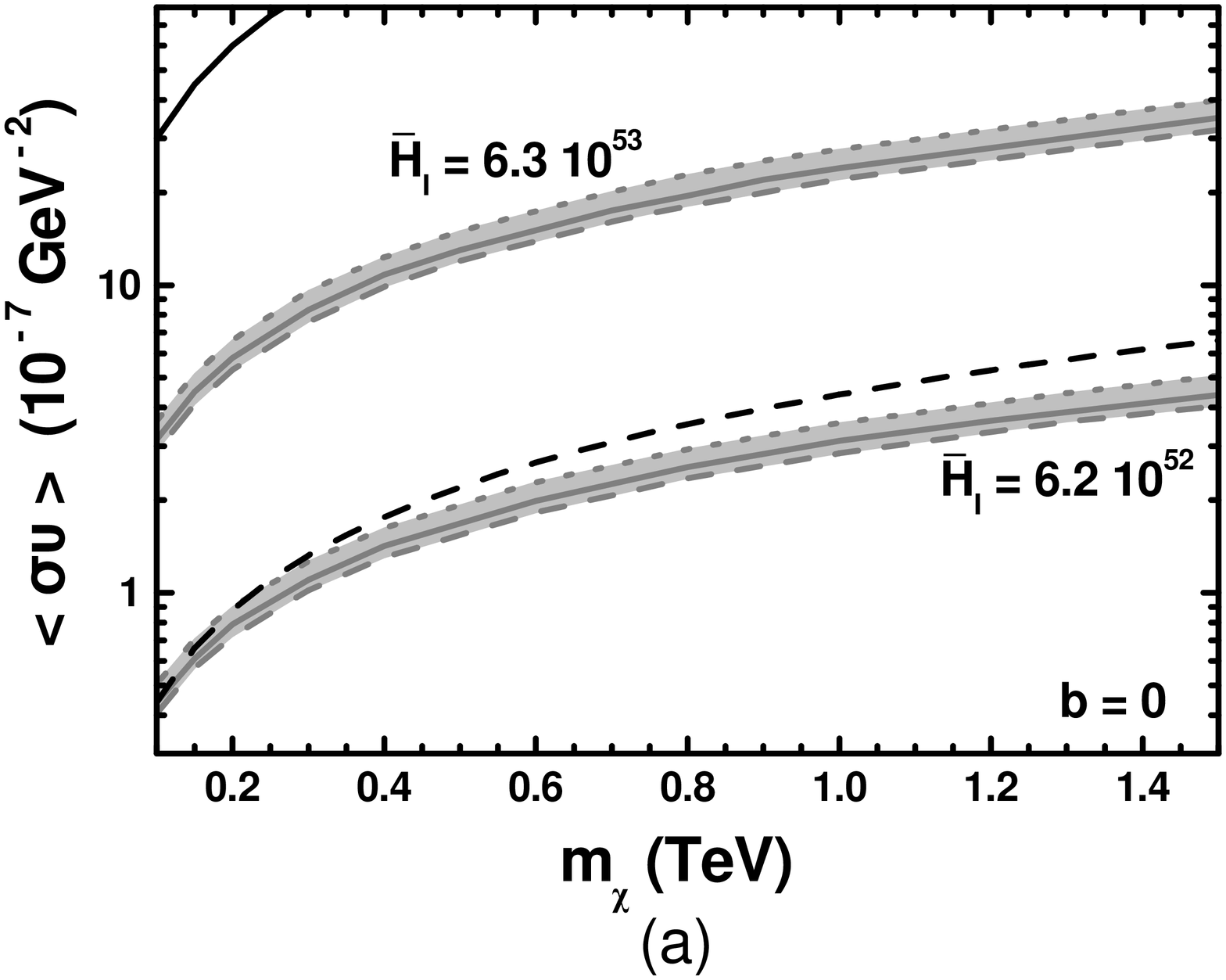,height=3.55in,angle=-90}
\hspace*{-1.37 cm}
\epsfig{file=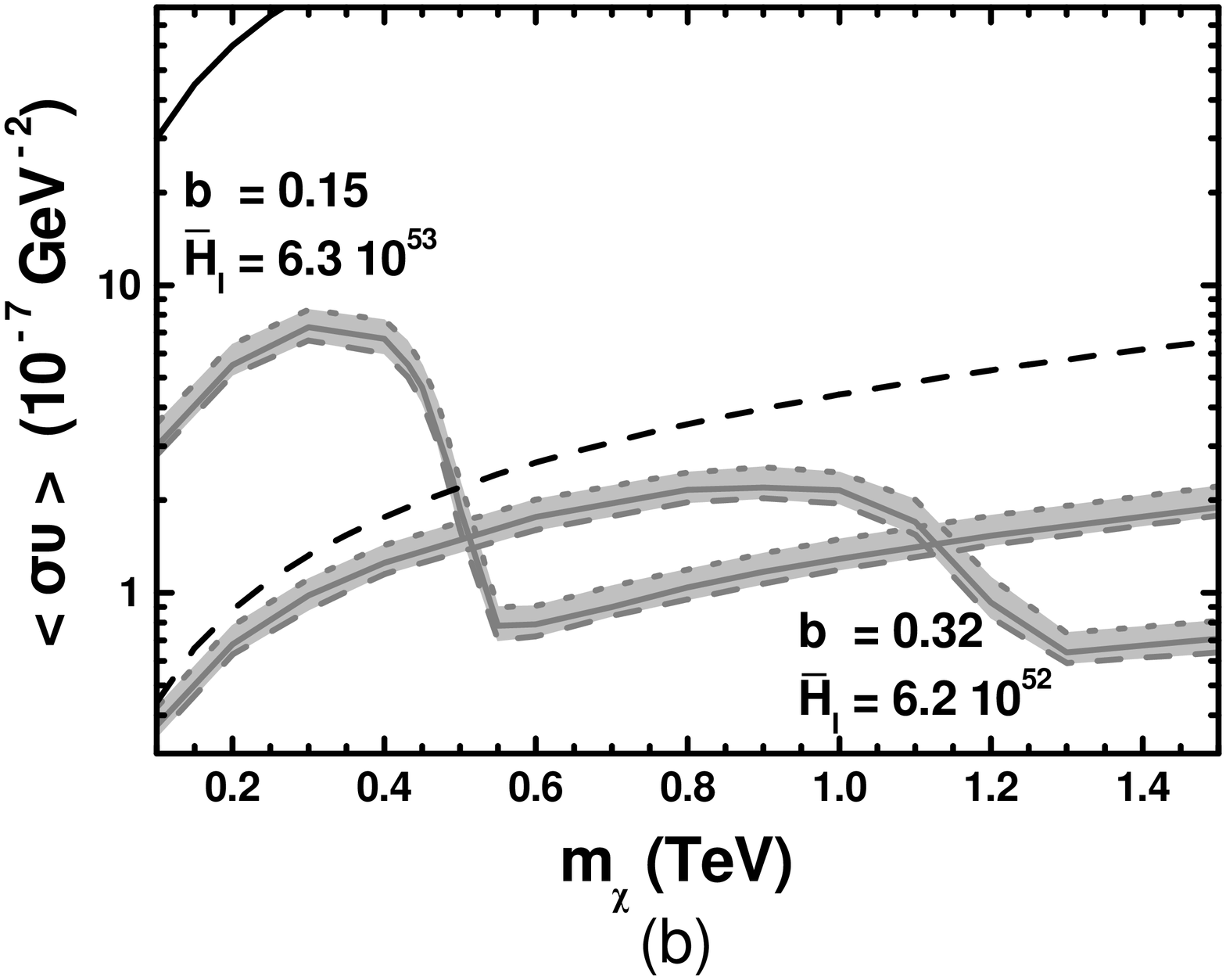,height=3.55in,angle=-90} \hfill
\end{minipage}\vspace*{-.05in}
\begin{minipage}{0.75\textwidth}
\captionof{figure}{\sl\ftn Restrictions in the $m_{\chi}-\sgv$
plane for $a=0.5,~\vTi=10^9~\GeV$ and several $b$'s and $\vHi$'s
indicated in the graphs {\sf\ssz (a)} and {\sf\ssz (b)}. The light
gray shaded areas are allowed by \Eref{cdmb} whereas the region
above the black solid [dashed] line is ruled out by the upper
bound on $\sgv$ from \sEref{sigBBN}{\sf\ssz a}
[\sEref{sigBBN}{\sf\ssz b}] assuming $\chi\chi\rightarrow\ps\el$.
The conventions adopted for the residual lines are also shown.}
\label{svmx}\hfill
\end{minipage}\hfill
\begin{minipage}{0.2\textwidth}
\vspace*{-.2in}\centerline{\epsfig{file=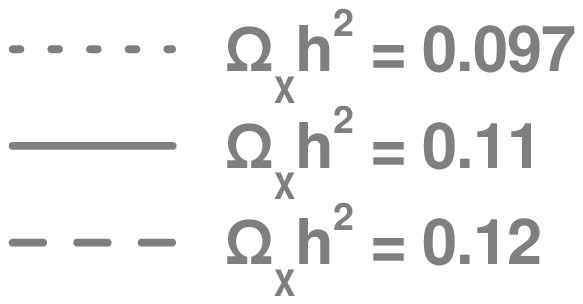,height=1.3in,angle=-90}}
\end{minipage}
\end{center}

In the plots of \Fref{svmx} the upper bounds from
\sEref{sigBBN}{a} and \sEref{sigBBN}{b} are denoted by black solid
(left corner of each plot) and dashed lines respectively. On the
other hand, the light gray shaded regions are confronted with
\Eref{cdmb}. The gray dashed [dotted] lines correspond to the
upper [lower] bound on $\Omx$ in \Eref{cdmb}, whereas the gray
solid lines are obtained by fixing $\Omx$ to its central value in
\Eref{cdmb}. We observe that $\Omx$ decreases as $\sgv$ increases.
This is due to the fact that $\Omx\propto1/\sgv$, as can be
deduced from \eqs{om1}{BEsol}. We also observe that for $\vtf$ far
away from $\vtp$ the region allowed by \Eref{cdmb} for $b\neq0$
reaches the one for $b=0$ with the same $\vHi$. However, when
$\vtf$ reaches $\vtp$, $\Omx$ decreases -- as we explain in
\Sref{sec:wimp} -- and therefore, the required for obtaining
$\Omx$ in the range of \Eref{cdmb} $\sgv$ decreases too. As a
consequence, although the allowed by \Eref{cdmb} area in
\sFref{svmx}{b} with $b=0.15$ approaches the corresponding area in
\sFref{svmx}{a} with the same $\vHi$ and violates the bound of
\sEref{sigBBN}{b} for low $\mx$'s, it becomes compatible with the
latter constraint for larger $\mx$'s.

We observe that, due to the dependence of $\Omx$ on the hierarchy
between $\vtf$ and $\vtp$ for $b\neq0$ we can achieve
compatibility of the CDM constraint with the bounds of
\sEref{sigBBN}{b} even for large $\Omqns$. On the contrary, this
can be attained for $b=0$ only tuning $\Omqns\lesssim10^{-4}$.
However in both cases ($b=0$ and $b\neq0$) agreement with the
requirement of \sEref{cdmb}{a} implies almost two-three orders of
magnitude higher $\sgv$'s than those required in the SC -- c.f.
\cref{jcapa}. It is worth mentioning that the obtained $\sgv$'s
can assist us to interpret \cite{khalil}, through WIMP
annihilation in the galaxy the reported excess on the positron
and/or electron cosmic-ray flux \cite{exper}, without invoking any
pole effect \cite{pole}, ad-hoc boost factor \cite{sommer} or
other astrophysical sources \cite{clumps}. According to
preliminary results \cite{patra}, the best fits to the
experimental data can be achieved for the annihilation channel
$\chi\chi\rightarrow \mu^+\mu^-$ with $\mx\simeq(1-1.6)~\TeV$ and
$\sgv\simeq(1-5)\cdot10^{-6}~\GeV^{-2}$. However, these values
remain tightly restricted by the CMB constraint \cite{CMB}.

\section{CDM from Thermal Production of \emph{e}-WIMPs}\label{sec:ewimp}

Similarly to the previous section, we analyze the behavior of
$\OmX$ as a function of the free parameters (\Sref{sec1:ewimp}),
and we identify the parametric regions allowed by the various
constraints (\Sref{sec2:ewimps}). As emphasized in
\Sref{BEewimps}, we focus on large $\Ti$'s which are frequently
met in the well-motivated models of SUSY inflation (see, e.g.,
\cref{susyhybrid}).

\subsection{$\Omega_X h^2$ as a Function of the Free Parameters}\label{sec1:ewimp}

\begin{figure}[!t]\vspace*{-.1in}
\hspace*{-.25in}
\begin{minipage}{8in}
\epsfig{file=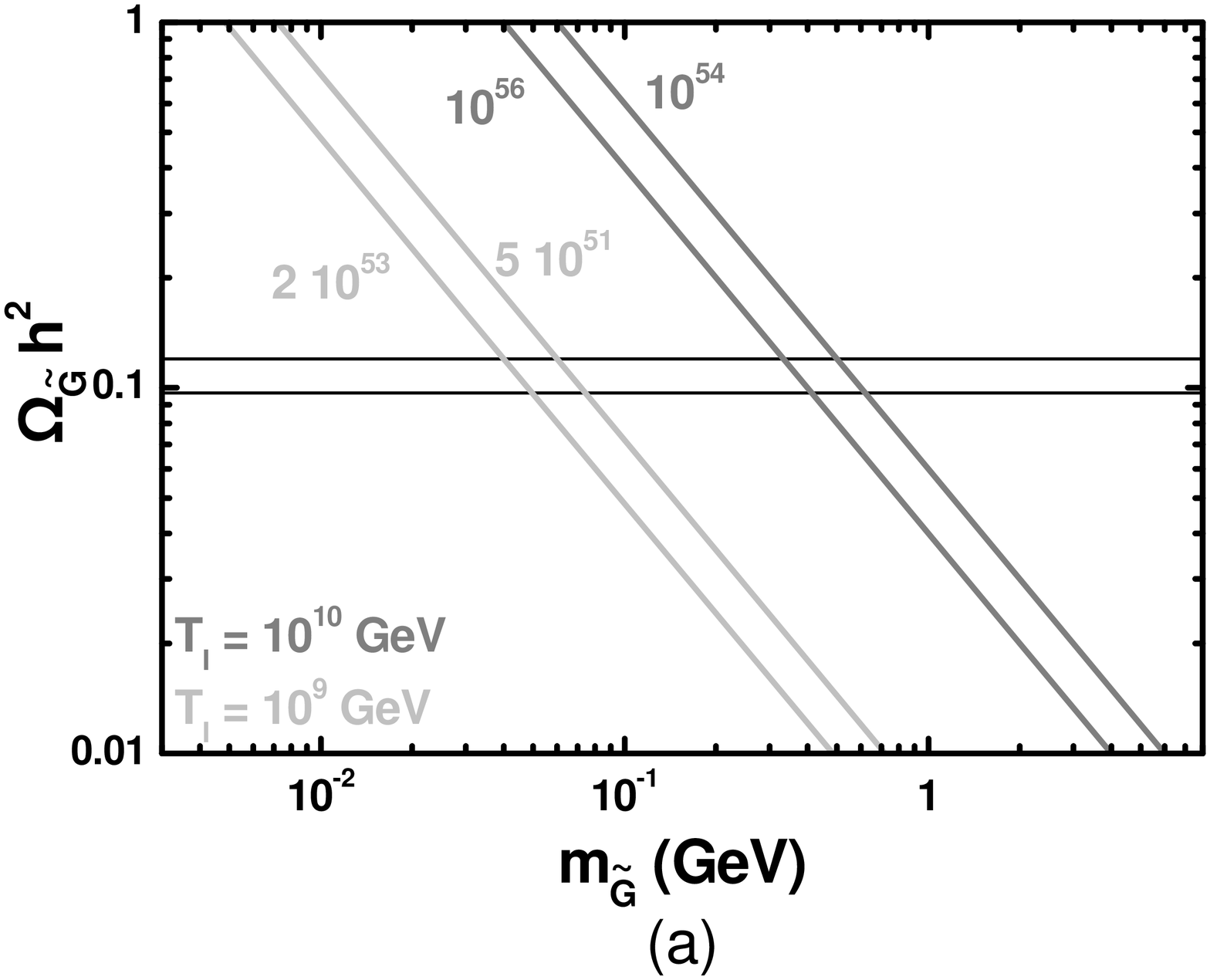,height=3.55in,angle=-90}
\hspace*{-1.37 cm}
\epsfig{file=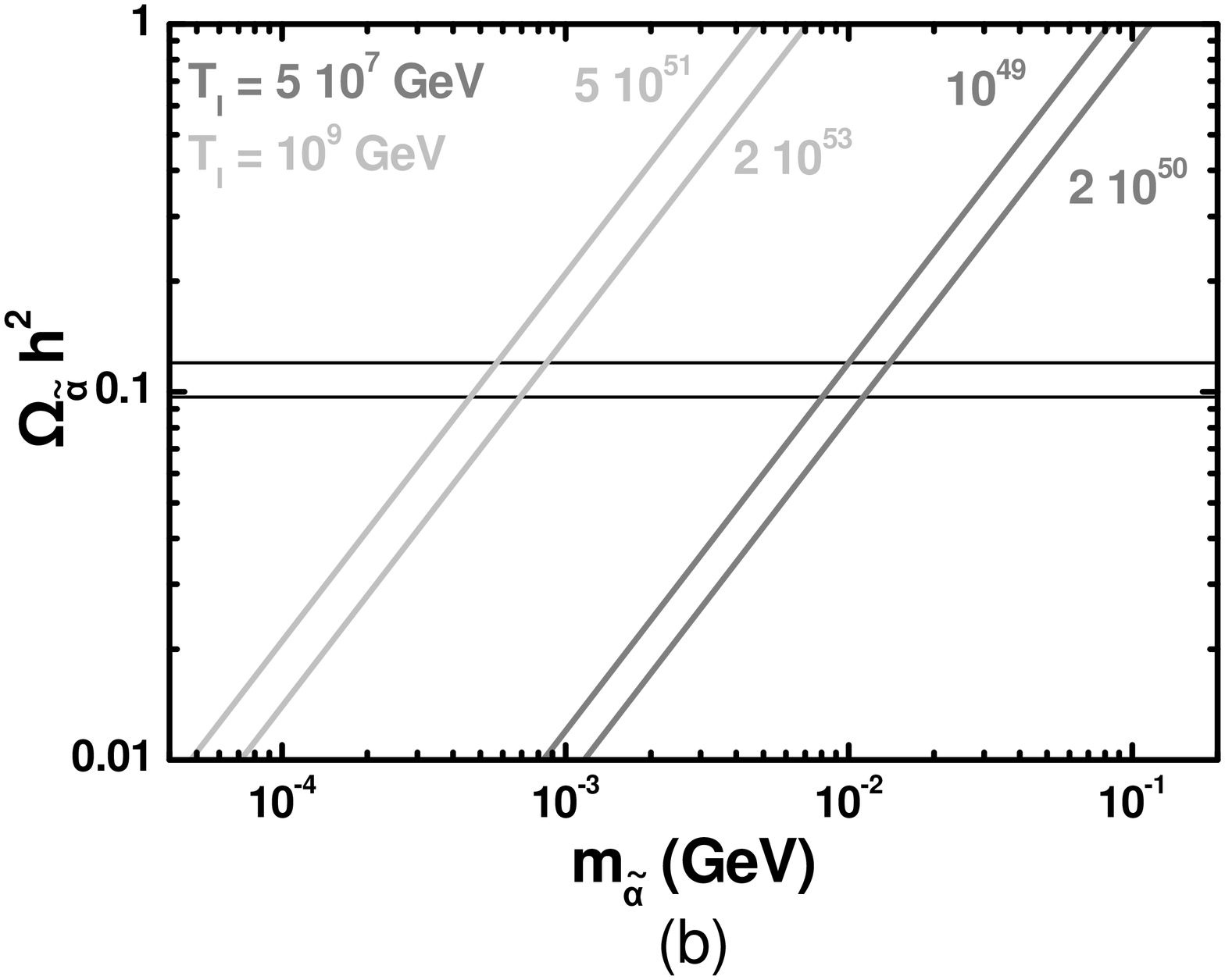,height=3.55in,angle=-90} \hfill
\end{minipage}
\hfill \caption[]{\sl\ftn $\Omega_{X}h^2$ as a function of $m_X$
($X=\Gr$ [$X=\ax$]) for various $\vHi$'s, indicated in the curves,
$a=0.5$, $b=0.2$ and $M_{1/2}=1~\TeV$ [$f_a=10^{11}~\GeV$] ({\ssz
\sf a} [{\ssz\sf b}]). We set {\ssz \sf (a)} $\Ti=10^9~\GeV$
(light gray lines) or $\Ti=10^{10}~\GeV$ (gray lines) and {\ssz
\sf (b)} $\Ti=10^9~\GeV$ (light gray lines) or
$\Ti=5\cdot10^{7}~\GeV$ (gray lines). The CDM bounds of
Eq.~(\ref{cdmb}) are, also, depicted by the two thin
lines.}\label{omX}
\end{figure}

By varying the free parameters, useful conclusions can be drawn
for the behavior of $\Omega_{\chi}h^2$.  The results are presented
in \sFref{omX}{a} [\sFref{omX}{b}] where we plot $\Omega_{X}h^2$
as a function of $m_X$ (with $X=\Gr$ [$X=\ax$]) for $a=0.5, b=0.2$
and $M_{1/2}=1~\TeV$ [$f_a=10^{11}~\GeV$], $\Ti=10^9~\GeV$ (light
gray lines), $\Ti=10^{10}~\GeV$ (gray lines) [$\Ti=10^9~\GeV$
(light gray lines) and $\Ti=5\cdot10^{7}~\GeV$ (gray lines)]. We
use various values for the $\vHi$, indicated in the curves. The
CDM bounds of Eq.~(\ref{cdmb}) are, also depicted by the two thin
lines. The results are derived numerically, but they can also be
reproduced from the empirical expressions in \Eref{emp}.

For each choice of $\Ti$ in \Fref{omX}, the selected $\vHi$'s
belong to the allowed region of \sFref{OmT}{b} and yield
$\Omqns>10^{-5}$. In all cases $\OmX$ is stabilized at $T\gg\Tc$
and therefore, its calculation exclusively depends on $C_X^{\rm
HT}$ given by \Eref{sig1}. For the $\Ti$'s and $\vHi$'s under
consideration, $\Omqns$ and the $\vtp$ closest to $\vti$, increase
with $\vHi$. Note that this fact is not explicitly shown in the
approximate \Eref{tmax}. As a consequence, $\OmX$ (which takes its
present value close to this $\vtp$) decreases as $\vHi$ increases.
This feature is similar to what happens in the pure KD era -- c.f.
\cref{huelva}. Similarly, we also observe $\Omgr\propto
1/m_{\Gr}$, whereas $\Omax\propto m_{\ax}$ -- see \Eref{emp}.

\subsection{Restrictions in the $m_X-\log\vHi$ Plane}\label{sec2:ewimps}

As already mentioned, $X$ constitutes a good CDM candidate if
$\OmX$ satisfies the criterion of Eq.~(\ref{cdmb}). Enforcing the
latter constraint we can derive restrictions in the $m_X-\log\vHi$
plane. We focus on $(\Ti,\vHi)$'s that belong in the first allowed
band -- depicted in \sFref{OmT}{b} -- of our QS. For the sake of
comparison, we present results even for $b=0$, although the
tracking behavior of our QS is not attained in this case as
explained in \Sref{Qev}.

\subsubsection{Gravitino Cold Dark Matter.}\label{sec:grv}

We recall that the free parameters in the present case are:
$m_{\Gr}, \Ti, M_{1/2}$ and $\vHi$ with fixed $b$ and $a=0.5$.

\begin{figure}[!t]\vspace*{-.1in}
\hspace*{-.3in}
\begin{minipage}{8in}
\epsfig{file=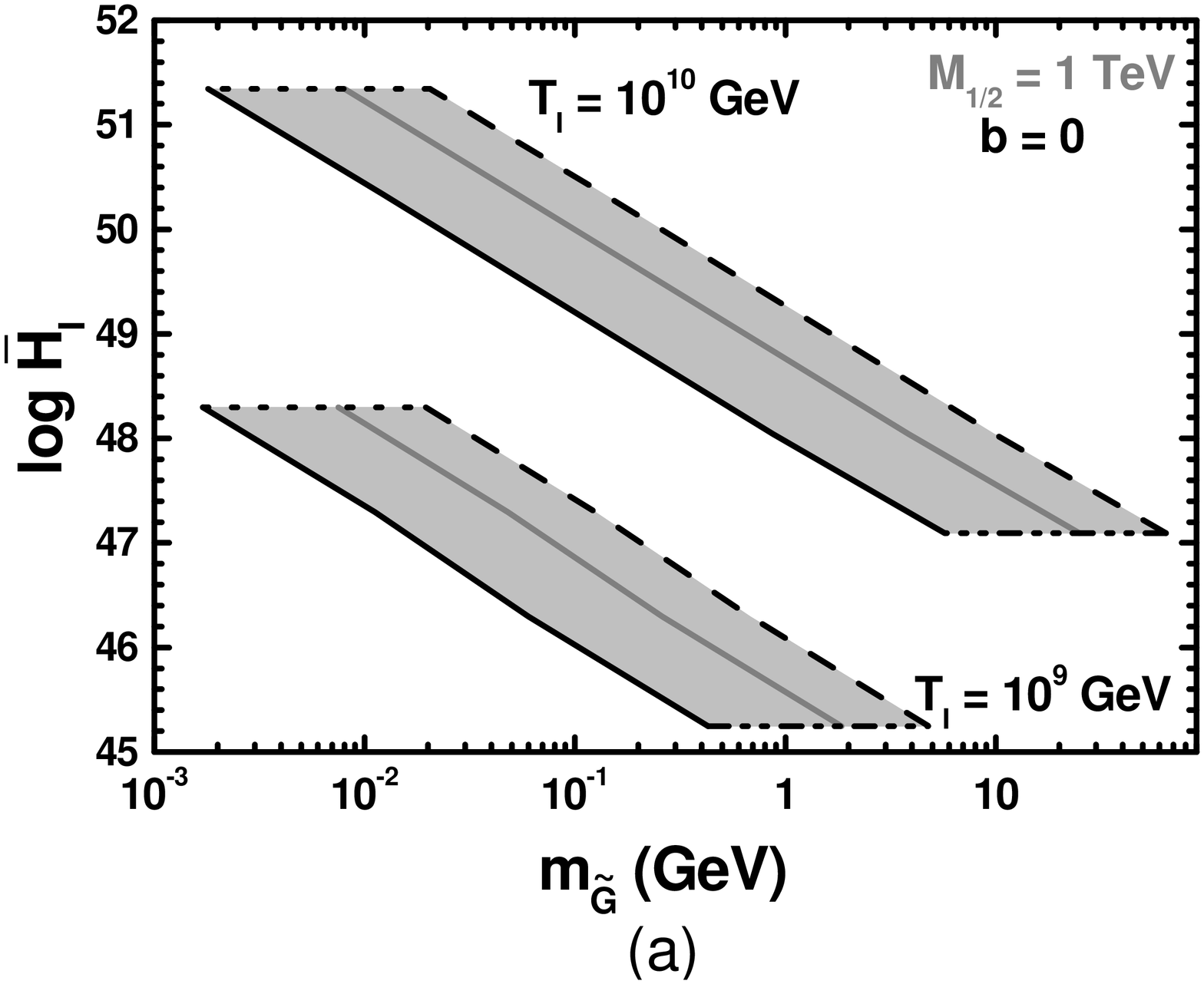,height=3.55in,angle=-90}
\hspace*{-1.37 cm}
\epsfig{file=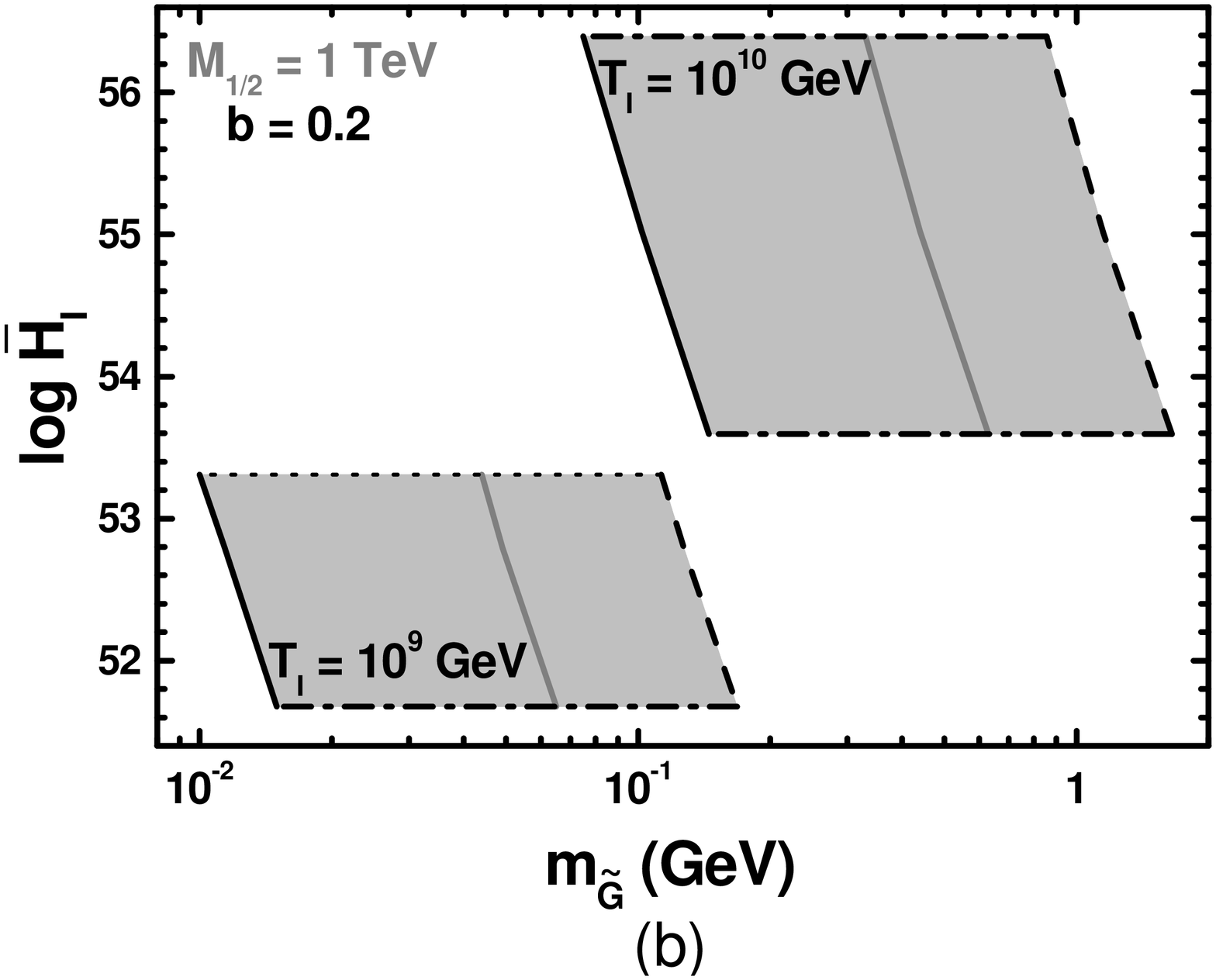,height=3.55in,angle=-90} \hfill
\end{minipage}\vspace*{-.1in}
\hfill\begin{center}
\epsfig{file=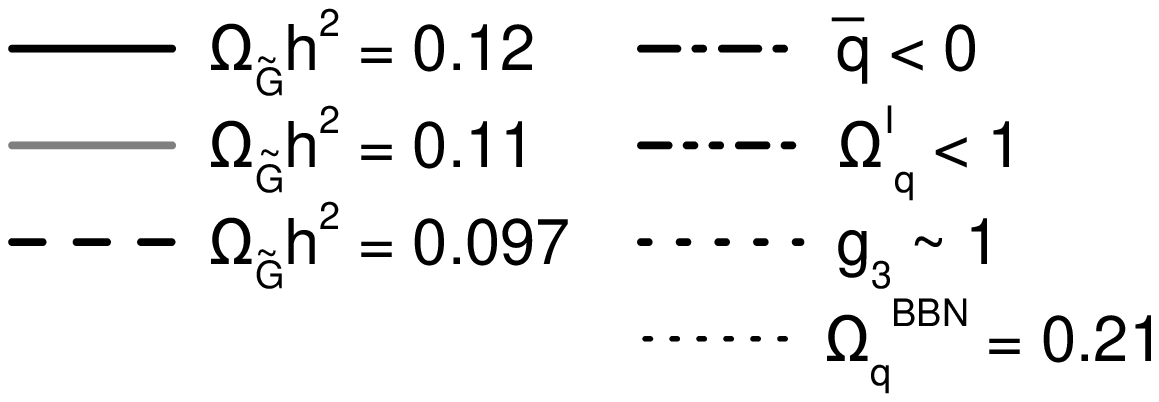,height=2.5in,angle=-90}
\end{center}
\hfill \caption[]{\sl\ftn Allowed (lightly gray shaded) regions in
the $m_{\Gr}-\log\vHi$ plane for $\Gr$ CDM with $0.5\leq
M_{1/2}/\TeV\leq1.5$, $a=0.5$, $\Ti=10^{10}~\GeV$ or
$\Ti=10^{9}~\GeV$ and {\sf\ssz (a)} $b=0$ and {\sf\ssz (b)}
$b=0.2$. The conventions adopted for the various lines are also
shown.}\label{Tmg}
\end{figure}

In \sFref{Tmg}{a} [\sFref{Tmg}{b}] we plot the allowed (lightly
gray shaded) regions in the $m_{\Gr}-\vHi$ plane, letting
$M_{1/2}$ vary in the interval $(0.5-1.5)~\TeV$, for $a=0.5$,
$b=0$ [$b=0.2$] and $\Ti=10^9~\GeV$ or $10^{10}~\GeV$. The black
solid [dashed] lines correspond to the upper [lower] bound on
$\Omgr$ in \sEref{cdmb}{a}, whereas the gray solid lines have been
obtained by fixing $\Omgr$ to its central value in \sEref{cdmb}{a}
for $M_{1/2}=1~\TeV$. Having in mind \Eref{emp}, we construct the
solid [dashed] line for $M_{1/2}=0.5~\TeV$ [$M_{1/2}=1.5~\TeV$].

The upper [lower] boundary curve (dotted [double dot-dashed] line)
of the allowed regions in \sFref{Tmg}{a} arises from the
saturation of $g_3<1$ [\Eref{domk}]. Recall that $g_3<1$ allows
employing $C_{\Gr}=C_{\Gr}^{\rm HT}$  self consistently in our
calculation. On the other hand, in \sFref{Tmg}{b}, the upper and
lower boundaries (dashed lines) of the allowed area for
$\Ti=10^{10}~\GeV$  arise from the band structure of our QS. This
is, also, the origin of the lower boundary (dashed line) of the
allowed area for $\Ti=10^9~\GeV$. The upper boundary (thin dotted
line) of this area comes from \Eref{nuc}. We observe that the
required $\mgr$'s increase with $\Ti$ as expected from \Eref{emp}.

As emphasized in \cref{huelva}, for $b=0$,
acceptable $\Omgr$'s require a fine
tuning of $\Omqns$'s to very low values. Indeed, in the allowed
regions of \sFref{Tmg}{a} for $\Ti=10^9~\GeV$ [$\Ti=10^{10}~\GeV$]
we have $10^{-19}\lesssim\Omqns\lesssim10^{-13}$
[$10^{-21}\lesssim\Omqns\lesssim10^{-13}$]. Such an unattractive
tuning is not needed for $b=0.2$. In fact, in the allowed regions of
\sFref{Tmg}{b} for $\Ti=10^9~\GeV$ [$\Ti=10^{10}~\GeV$] we have
$10^{-4}\lesssim\Omqns\lesssim0.21$
[$10^{-6}\lesssim\Omqns\lesssim0.064$]. We conclude, therefore, that
$\Gr$ is a natural CDM candidate within our QS.

\subsubsection{Axino Cold Dark Matter.}\label{sec:axn}

In considering the candidature of $\ax$ for the major CDM
component of the universe, we have initially to assume that the
scalar SUSY partner of $\ax$ (known as saxion) does not decay
\cite{axino} out-of-equilibrium producing entropy and thereby,
diluting $\Omax$. We then have to ensure the consistency of the
hypothesis that $\ax$ decouples from the thermal bath at a
temperature $T_{\rm D}>\Ti$. To this end, we check that for every
$T<\Ti$ the following condition is valid:
\beq \label{TD}
H(T)>\Gamma_{\ax}(T)~~\mbox{where}~~\Gamma_{\ax}\sim 6\,N_{\rm
F}(N_3^2-1)g_a^2g_3^2\, n^{\rm eq}/2. \eeq
Here, $\Gamma_{\ax}$
is the interaction rate of $\ax$'s with
the thermal bath \cite{axinold},
$N_{\rm F}=12$ and $N_3=3$ \cite{huelva}. The free parameters in the present
case are: $m_{\ax},~f_{a},~\Ti$ and $\vHi$ with fixed $b$ and
$a=0.5$.

\begin{figure}[!t]\vspace*{-.1in}
\hspace*{-.3in}
\begin{minipage}{8in}
\epsfig{file=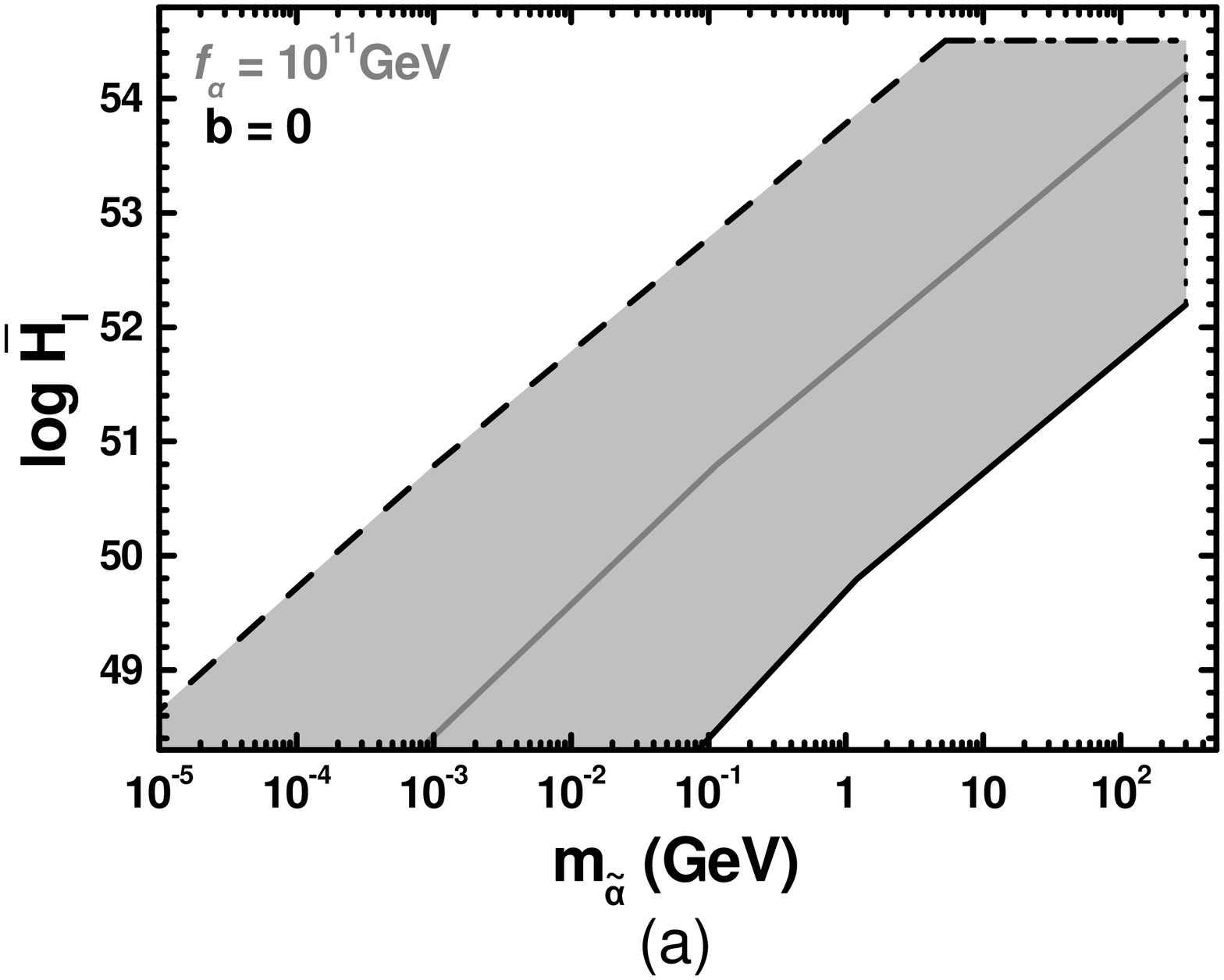,height=3.55in,angle=-90}
\hspace*{-1.37 cm}
\epsfig{file=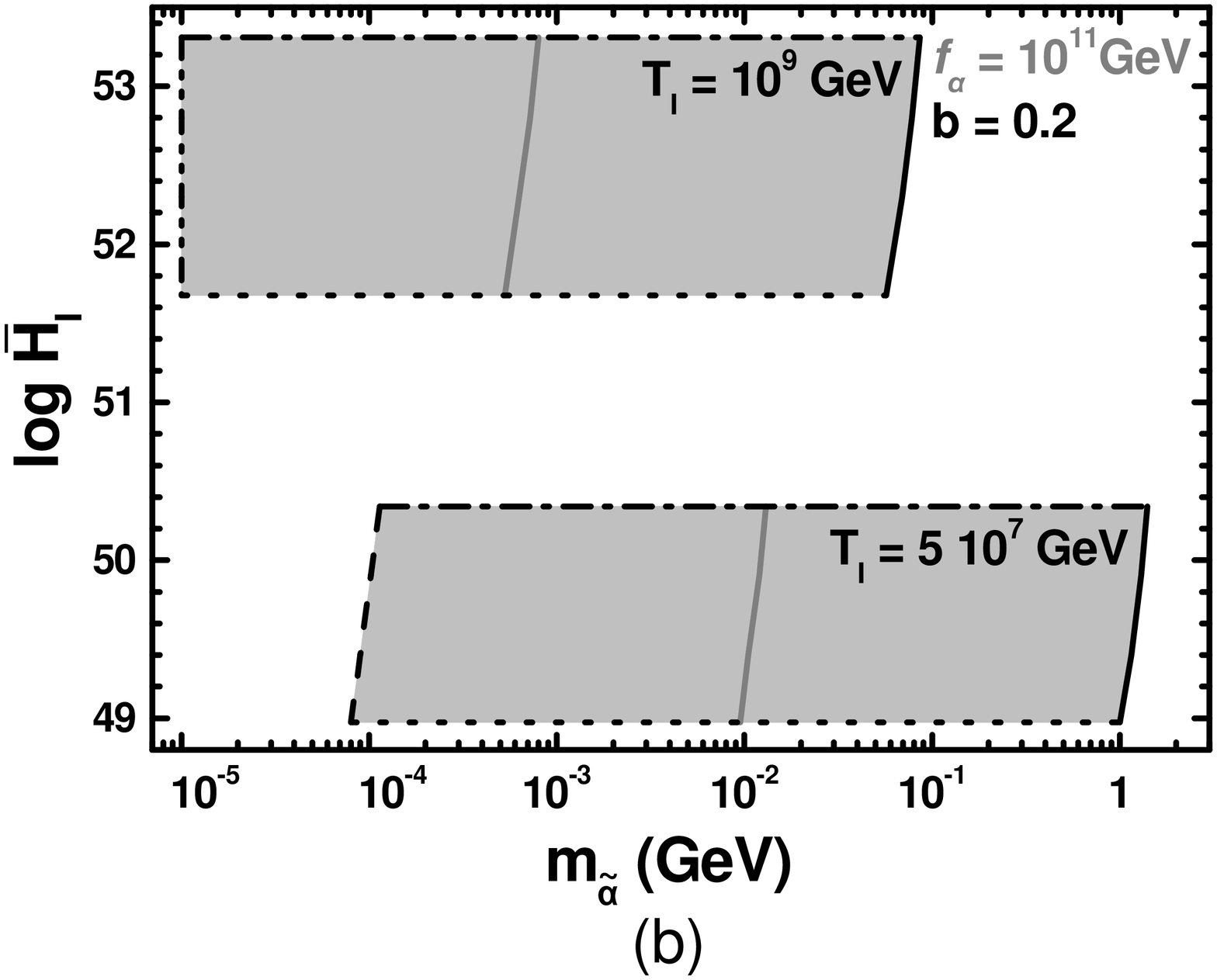,height=3.55in,angle=-90} \hfill
\end{minipage}\vspace*{-.1in}
\begin{center}
\epsfig{file=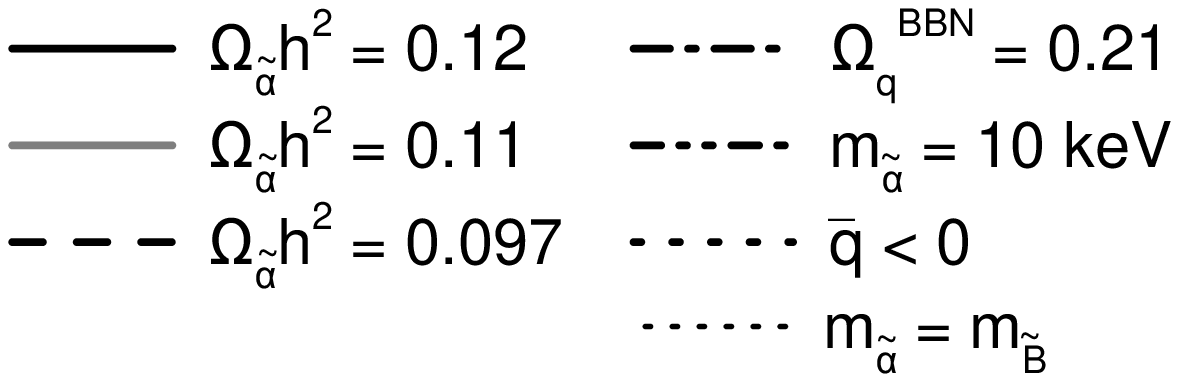,height=2.5in,angle=-90}
\end{center}
\hfill \caption[]{\sl\ftn Allowed (lightly gray shaded) regions in
the $m_{\ax}-\log\vHi$ plane for $\ax$ CDM, $10^{10}\leq
f_a/\GeV\leq10^{12}$, $a=0.5$, $\Ti=10^9~\GeV$ or $\Ti=5\cdot
10^7~\GeV$ and {\sf\ssz (a)} $b=0$ or {\sf\ssz (b)} $b=0.2$. For
$b=0$ we use the $m_i$'s shown in \Eref{mi}. The conventions
adopted for the various lines are also shown.}\label{Tmd}
\end{figure}

In \sFref{Tmd}{a} [\sFref{Tmd}{b}] we display the allowed (lightly
gray shaded) regions in the $m_{\ax}-\vHi$ plane for $10^{10}\leq
f_a/\GeV\leq10^{12}~$, $a=0.5$, $b=0$ [$b=0.2$] and
$\Ti=10^9~\GeV$ or $\Ti=5\cdot10^7~\GeV$. The black solid [dashed]
lines correspond to the upper [lower] bound on $\Omax$ in
\sEref{cdmb}{a}, whereas the gray solid lines have been obtained
by fixing $\Omax$ to its central value in \sEref{cdmb}{a} for
$f_a=10^{11}~\GeV$. In practice, the solid [dashed] line is
constructed for $f_a=10^{12}~\GeV$ [$f_a=10^{10}~\GeV$]. This can
be understood taking into account the empirical relations for
$\Omax$ given in \cref{huelva} [\Eref{emp}] for $b=0$ [$b=0.2$].

The upper boundary curves (dot-dashed line) of the allowed areas
in \Fref{Tmd} come from the upper bound on $\Omqns$ in \Eref{nuc}.
The right boundary (thin line) of the allowed area in
\sFref{Tmd}{a} arises from the upper bound of \sEref{cdmb}{b}
assuming that $\tilde B$ is the NLSP, with a mass as in \Eref{mi}.
Needless to say that this upper bound can be modified if there is
another SUSY particle lighter than $\tilde B$. The relevant area
terminates from below at $\Tkr\simeq 1~\TeV$, so that our formulas
for $C^{\rm LT}_{\ax}$ in \cref{huelva} are fully applicable. The
low boundary curves of the allowed areas in \sFref{Tmd}{b} arise
from the band structure of the parameter space of the  QS under
study.

Contrary to the case of $\Gr$, we observe that the required
$\mxx$'s increase when $\Ti$ decreases, as expected from
\Eref{emp}. In the allowed region of \sFref{Tmd}{a} we get
$10^{-13}\lesssim\Omqns\lesssim0.21$. In this case (with $b=0$),
the $\Omax$ calculation is realized employing $C^{\rm LT}_{\ax}$
corresponding to the $m_i$'s indicated in \Eref{mi}. As outlined
in \cref{huelva} and deduced from \sFref{Ys}{b}, the $\Omax$
calculation in this regime -- and therefore, the allowed area of
\sFref{Tmd}{a} -- is independent of $\Ti$, provided $\Ti>\Ts$. On
the other hand, in the allowed regions of \sFref{Tmd}{b} for
$\Ti=10^9~\GeV$ [$\Ti=5\cdot10^{7}~\GeV$] (and $b=0.2$), we have
$10^{-4}\lesssim\Omqns\lesssim0.21$
[$0.031\lesssim\Omqns\lesssim0.21$] and the $\Omax$ calculation
depends exclusively on $C^{\rm HT}_{\ax}$, as underlined in
\Sref{BEewimps}. Larger $\mxx$'s are allowed in the case of
\sFref{Tmd}{a}. Obviously, in both cases $\ax$ turns out to be a
natural CDM candidate for a wide range of $\mxx$'s. However,
within our QS ($b>0$), this result is insensitive to the low
energy s-particle spectrum of the theory but depends on $\Ti$.

\section{Conclusions}\label{sec:con}

We studied a quintessential model  based on an inverse-power-law
potential supplemented with a Hubble-induced mass term for the
quintessence field, $q$ -- see \Eref{qeq}. We verified that this
term ensures the presence of a period dominated by the kinetic
energy of $q$ and allows the quintessential energy density to
develop a tracker behavior sufficiently early, alleviating in this
way the coincidence problem. In addition to the numerical
treatment of the relevant equations (which is mandatory in order
to obtain a reliable description of the quintessential dynamics)
we presented a qualitative but rather comprehensive
semi-analytical approach. The parameters of the model ($a, b, \Ti,
\vHi$) were confined so as $\Omega_q(\Ti)=1$ and were constrained
by current observational data originating from BBN, the present
acceleration of the universe, the inflationary scale and the DE
density parameter.

We found that $0<a<0.6$ and that there is a reasonably
allowed region in the ($b, \vHi$) plane with $b$ mildly tuned to
values of order $0.1$. Extrapolating the results of \cref{mas} to
higher temperatures, we showed that, contrary to the pure KD
era, the KD generated in this model is characterized by an
oscillatory evolution of $q$ and the barotropic index.

We then examined the impact of this modified KD epoch on the
thermal abundance, $\OmX$, of WIMPs  and \emph{e}-WIMPs. Solving
the problem numerically and semi-analytically we found that:

\begin{itemize}
\item $\Omx$, with $\chi$ a WIMP, increases w.r.t its value in the
SC. Its increase is not monotonic as in the case of a pure KD era,
but crucially depends on the hierarchy between the freeze-out
temperature and the temperature where the evolution of $q$
develops extrema.

\item $\OmX$ with $X$ an \emph{e}-WIMP (gravitino, $\Gr$,  or
axino, $\ax$) takes its present value at the closest temperature
to $\Ti$, where $q$ develops its extremum. As a consequence, while
$\OmX$ decreases w.r.t its value in SC, it  increases w.r.t its
value in the pure KD phase, and both $\Gr$ and $\ax$ arise as natural
CDM candidates for masses in the range $(10^{-4}-1)~\GeV$.
\end{itemize}

We note that, additional BBN bounds might arise due to possible
decays of the NLSP. Including these effects would have introduced
significant model-dependence in our analysis, and would be
relevant mainly for gravitinos, whose interactions are extremently
suppressed. Even in this case the changes would be in at the
numerical level, while the qualitative features of our discussion
remain valid.

Given that we did not adopt a specific theoretical framework in
our approach, we kept for simplicity and better definiteness $b$
constant during the various phases of the $q$ evolution -- c.f.
\cref{mas}. However, within supergravity, $b$ may change from
phase to phase -- see, e.g., \cref{moroim} --, thus affecting the
quintessential dynamics -- see \Sref{Qev}. We checked that the
value of $b$ is to remain almost constant (at the level of
$\pm10\%$) during the RD era which follows the KD era, for the
tracker solution to be joined in time. On the contrary, if we
switch off $V_b$ after the onset of the matter domination, our
results on $\Omega_{q0}$ and $w_q(0)$ remain more or less intact.
Moreover, $\OmX$, when $X$ is a WIMP, increases [decreases] when
$b$ decreases [increases] after the KD era -- see \Eref{BEsol}. On
the other hand, $\OmX$, if $X$ is an \emph{e}-WIMP, is not so
sensitive to possible alterations of $b$ during the post-kination
phases, since its magnitude is determined mainly during the KD
era.

Further work \cite{patra} is required in order to establish
whether the enhancements of $\Omx$ obtained in our QS can explain
the reported \cite{exper} results on the cosmic-ray fluxes through
WIMP annihilation in the galaxy. In addition, it would be
interesting to check whether the extrema in the evolution of $q$
affect the interference between thermal leptogenesis and neutrino
masses in conjunction with the $\Gr$ constraint -- see, e.g.,
Ref.~\cite{kinlept}.

\acknowledgments S.L and C.P have been supported by the FP6 Marie
Curie Excellence grant MEXT-CT-2004-014297. SL also acknowledges
support by the European Research and Training Network UniverseNet,
MRTPN-CT-2006 035863-1.

\appendix{Tracking Quintessence and the $\Gr$ Constraint}

\rhead[\fancyplain{}{ \bf \thepage}]{\fancyplain{}{\scshape
Tracking Quintessence and CDM Candidates}}
\lhead[\fancyplain{}{\sc Tracking Quintessence and the $\Gr$
Constraint}]{\fancyplain{}{\bf \thepage}} \cfoot{}

In this appendix, we analyze the implications of our
quintessential scheme for the unstable $\Gr$. In this case, $\Gr$
can decay after the onset of BBN, affecting the primordial
abundances of the light elements in an unacceptable way. In order
to avoid spoiling the success of BBN, an upper bound on $Y_{\Gr}$
is to be extracted as a function of $\mgr$ and the hadronic
branching ratio of $\Gr$, $B_{\rm h}$ \cite{kohri, kohri2,
oliveg}. This is the well-known $\Gr$ constraint. In what follows,
we specify some representative values of this constraint, taking
into account the most up-to-date analysis of Ref.~\cite{kohri}.
Note that we here review -- c.f. \cref{huelva} -- the $\mgr$'s
which correspond to different $Y_{\Gr}(\Tns)$'s, decoding more
precisely the relevant figures. In particular, if $\Gr$ decays
mainly to photon and photino, from Fig. 1 of Ref.~\cite{kohri} we
can deduce:
\beq \label{grph} Y_{\Gr}(\Tns)\lesssim\left\{\matrix{
10^{-15} \cr
10^{-14}\cr
10^{-13} \cr}
\right.~~\mbox{for}~~ \mgr\simeq \left\{\matrix{
0.43~{\rm TeV} \cr
0.69~{\rm TeV} \cr
10.6~{\rm TeV} \cr}\right.~~\mbox{and}~~B_{\rm h}=0.001, \eeq
whereas if $\Gr$ decays mainly to gluons and gluinos, from Fig. 2
of Ref.~\cite{kohri} we can deduce:
\beq \label{grgl} Y_{\Gr}(\Tns)\lesssim\left\{\matrix{
10^{-15} \cr
10^{-16} \cr
8.5\cdot10^{-15} \cr}
\right.~~\mbox{for}~~ \mgr\simeq \left\{\matrix{
0.2~{\rm TeV}\hfill \cr
0.67~{\rm TeV}\hfill \cr
10~{\rm TeV} \hfill \cr}\right.~~\mbox{and}~~B_{\rm h}=1.\eeq
In the SC (where no late-time entropy production is expected)
\eqs{grph}{grgl} imply stringent upper bounds on $\Ti$, for fixed
$M_{1/2}$. For the indicative value $M_{1/2}=1~\TeV$, we find
\begin{equation} \label{bTr} T_{\rm I}\lesssim\left\{\matrix{
10^{6}~{\rm GeV}\hfill\cr
2.3\cdot10^{7}~{\rm GeV}\hfill \cr
5.6\cdot10^{8}~{\rm GeV}\hfill \cr}
\right. \mbox{for}~~ m_{\Gr}\simeq \left\{\matrix{
0.43~{\rm TeV} \cr
0.69~{\rm TeV} \cr
10.6~{\rm TeV} \cr}\right.~~\mbox{and}~~B_{\rm
h}=0.001,~~\mbox{or}\end{equation}
\begin{equation} \label{bTrh}  T_{\rm I}\lesssim\left\{\matrix{
1.45\cdot10^{5}~{\rm GeV}\hfill\cr
9\cdot10^{4}~{\rm GeV}\hfill \cr
4.8\cdot10^{7}~{\rm GeV}\hfill \cr}
\right. \mbox{for}~~ m_{\Gr}\simeq \left\{\matrix{
0.2~{\rm TeV} \cr
0.67~{\rm TeV} \cr
10~{\rm TeV} \cr}\right.~~\mbox{and}~~B_{\rm h}=1.\end{equation}
Clearly, the upper bound on $T_{\rm I}$ becomes significantly more
restrictive for large $B_{\rm h}$'s and low $m_{\Gr}$'s. These
restrictions on $T_{\rm I}$ can be avoided in both the pure
kination scenario and our QS.

\renewcommand{\arraystretch}{1.1}
\begin{table}[!t]\begin{tabular}[!t]{cc}\begin{minipage}[t]{5cm}
\begin{tabular}{|c||c|c|} \hline $\mgr$&$\vHi^{\rm
min}$&$\Tkr^{\rm max}$\\
$(\TeV)$&&$(\GeV)$\\\hline\hline
&\multicolumn{2}{|c|}{$B_{\rm h}=0.001$}\\\cline{2-3}
$0.43$&$1.3\cdot10^{46}$&$1.1\cdot10^5$\\
$0.69$&$3.5\cdot10^{44}$&$3.9\cdot10^6$\\
$10.6$&$5\cdot10^{42}$&$2.7\cdot10^8$\\\hline
&\multicolumn{2}{|c|}{$B_{\rm h}=1$}\\\cline{2-3}
$0.2$&$7.9\cdot10^{46}$&$2.5\cdot10^4$\\
$0.67$&$1.1\cdot10^{47}$&$1.7\cdot10^4$\\
$10$&$1.5\cdot10^{44}$&$8.7\cdot10^6$\\\hline
\end{tabular}\\\bec \vspace*{0.cm}  {\sf\small (a)}\eec\end{minipage}
&\begin{minipage}[t]{10cm}\hspace{0.4in}{\begin{tabular}{|c||c|c||c|c|}
\hline
$\mgr$&\multicolumn{2}{|c||}{$b=0.15$}&\multicolumn{2}{|c|}{$b=0.32$}\\\cline{2-5}
$(\TeV)$&\multicolumn{4}{|c|}{$\vHi/10^{50}$}\\\cline{2-5}
&$3.9\cdot10^3$&$1.8\cdot10^4$&$1.7\cdot10^2$&$7.9\cdot10^{3}$\\\cline{2-5}
&\multicolumn{4}{|c|}{$\Ygr/10^{-17}$}
\\\hline \hline
$0.43$&$8.7$&$7.6$&$22$&$15$\\
$0.69$&$4.6$&$4$&$12$&$8.2$\\
$10.6$&$1.96$&$1.7$&$5$&$3.6$\\
\hline
$0.2$&$33$&$29$&$85$&$59$\\
$0.67$&$4.7$&$4$&$12$&$8.7$\\
$10$&$1.96$&$1.7$&$5$&$3.6$\\
\hline
\end{tabular}\\ \bec \vspace*{0.0cm}  {\sf\small (b)}\eec} \hfill
\end{minipage}
\end{tabular}
\hfill \vspace*{-.3in}\caption[]{\sl \ftn The minimum values of
$\vHi,~\vHi^{\rm min}$, and the maximum values of $\Tkr,~\Tkr^{\rm
max}$, dictated by the $\Gr$ constraint for $b=0$ and $B_{\rm
h}=0.001$ or 1 {\sf\ssz (a)} and the obtained $\Ygr(\Tns)$ for
$b=0.15$ and $b=0.32$ and the boundary values of $\vHi$'s of the
first allowed band depicted in \sFref{OmT}{\ssz a} {\sf\ssz (b)}.
We use various $\mgr$'s indicated in the tables, $a=0.5$,
$\vTi=10^9~\GeV$ and $M_{1/2}=1~\TeV$.}\label{grtable}
\end{table}
\renewcommand{\arraystretch}{1.0}

Indeed, in the case of a pure KD era the $\Gr$ constraint entails
a lower bound on $\vHi$, $\vHi^{\rm min}$, which can be
transformed to an upper bound on $\Tkr$, $\Tkr^{\rm max}$, for
fixed $\Ti$, $M_{1/2}$ and $\mgr$ -- c.f. \cref{huelva}. Setting
$\Ti=10^9~\GeV$ and $M_{1/2}=1~\TeV$, we present in
Table~\ref{grtable}-{\sf\small (a)} the corresponding $\vHi^{\rm
min}$'s and $\Tkr^{\rm max}$'s for several $\mgr$'s and $B_{\rm
h}$'s. Clearly, as the $\Ygr$'s decrease, the required $\vHi^{\rm
min}$'s [$\Tkr^{\rm max}$'s] increase [decrease]. On the other
hand, as we emphasized in \Sref{BEewimps}, $\Ygr$ within our QS is
stabilized close to the temperature $T_{\rm exp}(k=0)$ which
correspond to $\vtp(k=0)$ -- see \Eref{tmax} -- where the earliest
peak of the $q$ evolution occurs. Therefore, we expect that the
$\Gr$ constraint imposes an upper bound on $T_{\rm exp}(k=0)$,
which is a function of $H_{\rm I}$ and $\Ti$. However, due to the
band structure of the allowed parameter space of our QS, only
certain $\vHi$'s are available for each $b$ -- see \sFref{OmT}{a}.
For this reason, we opt to present in
Table~\ref{grtable}-{\sf\small (b)}  the $\Ygr$'s resulting on the
boundaries of the first allowed band for $b=0.15$ and $b=0.32$
fixing again $\Ti=10^9~\GeV$ and $M_{1/2}=1~\TeV$. For the
selected $\vHi$'s we obtain $T_{\rm
ext}(k=0)\simeq2.5\cdot10^7~\GeV$ [$T_{\rm
ext}(k=0)\simeq8\cdot10^7~\GeV$] for $b=0.15$ [$b=0.32$]. We
observe that the obtained $\Ygr$'s decrease with increasing $\vHi$
(as for $b=0$) and are well below the bounds of \eqs{grph}{grgl}
besides for $b=0.32$, $\mgr=0.67~\TeV$ and $B_{\rm h}=1$ where
$\Ygr$ marginally violates the relevant bound. As a consequence,
the $\Gr$ constraint can be comfortably eluded for the
$(b,\vHi)$'s used in Fig.~\ref{om} and \ref{svmx} since the
employed there $\vHi$'s belong in the ranges of $\vHi$ examined in
Table~\ref{grtable}-{\small\sf (b)}. Comparing the results of
Table~\ref{grtable}-{\small\sf (a)} and {\small\sf (b)} we remark
that evading the $\Gr$ constraint requires larger $\vHi$'s for
$b\neq0$ than for $b=0$ -- see, e.g., the entries of two tables
for $\mgr=0.67~\TeV$ and $B_{\rm h}=1$.

\begin{figure}[t]\vspace*{-.3in}
\begin{center}
\epsfig{file=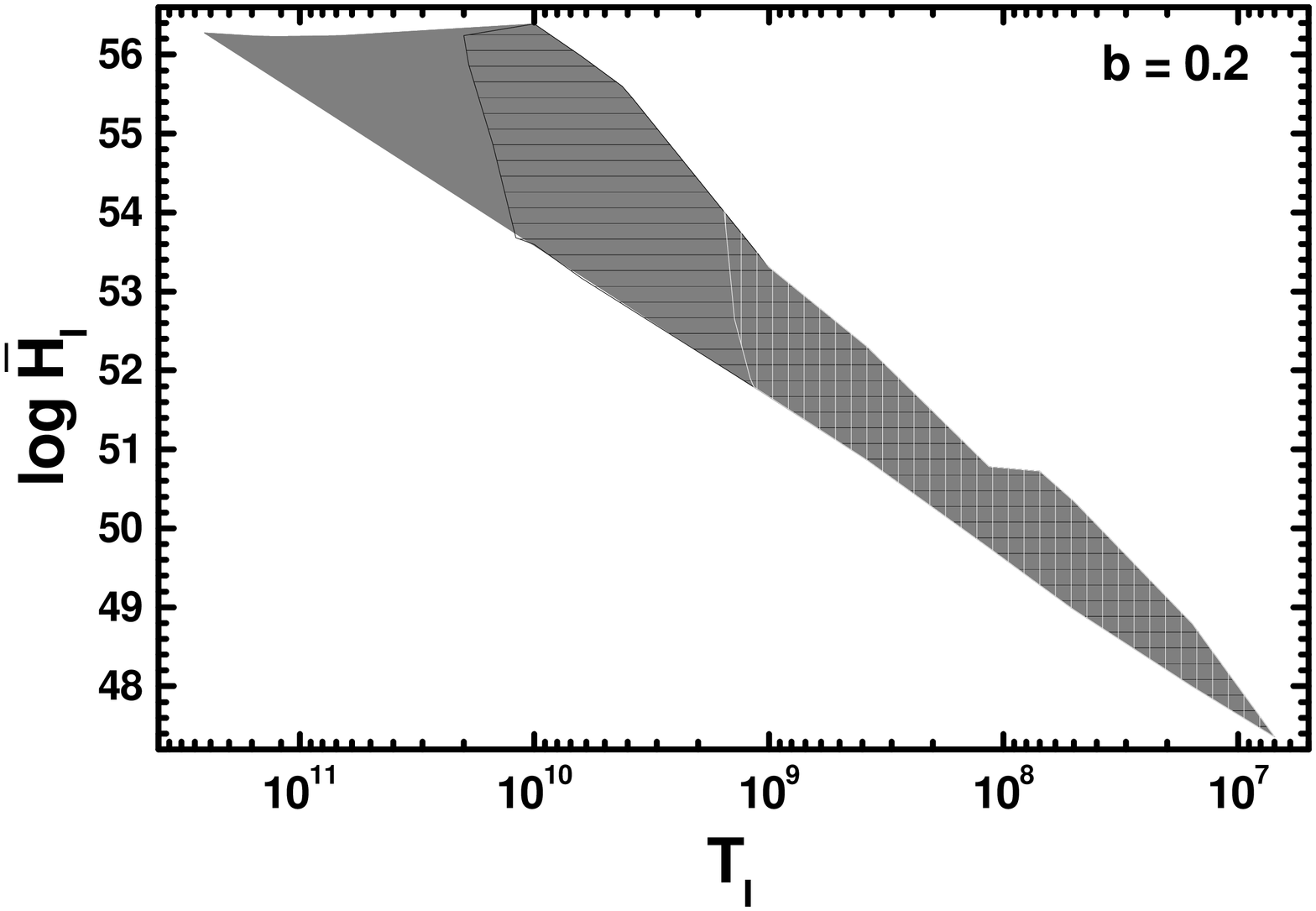,height=3.55in,angle=-90}
\end{center}
\hfill \vspace*{-.13in} \caption[]{\sl \ftn  Regions in the
$\Ti-\log\vHi$ plane that are simultaneously allowed by the
quintessential requirements (gray shaded area) --
Eqs.~(\ref{domk})-(\ref{wqd}) -- and the $\Gr$ constraint for
$M_{1/2}=1~{\rm TeV}$, $\mgr=0.5~{\rm TeV}$ and $B_{\rm h}=0.001$
(black lined area) or $B_{\rm h}=1$ (white lined area). We take
$a=0.5$, $b=0.2$ and $\vqi=0.01$.} \label{OmTgr}
\end{figure}

The importance of the KD era in weakening the $\Gr$ constraint
within our QS can be also induced by Fig.~\ref{OmTgr}, where, in
contrast to our previous approach, $\Ti$ is variable, whereas
$m_{\Gr}$ is fixed to a representative value. Namely, in
Fig.~\ref{OmTgr}, we show the regions in the $\log\Ti-\log\Omqns$
plane that are allowed by the quintessential requirements -- see
\sFref{OmT}{b} --, for $m_{\Gr}=0.5~\TeV$, $M_{1/2}=1~\TeV$ and
$B_{\rm h}=0.001$ (black lined area) or $B_{\rm h}=1$ (white lined
area). We observe that for $B_{\rm h}=0.001$ the allowed maximal
$\Ti$ is higher that in the case of $B_{\rm h}=1$. This is because
for $B_{\rm h}=0.001$ we impose
$Y_{\Gr}(\Tns)\lesssim1.7\cdot10^{-15}$, whereas for $B_{\rm
h}=1$, we impose $Y_{\Gr}(\Tns)\lesssim1.7\cdot10^{-16}$ (in
accordance with Figs 1 and 2 of Ref.~\cite{kohri}). As a
consequence, the maximal allowed $T_{\rm ext}(k=0)\simeq
(4.8\cdot10^5-6.8\cdot10^6)~\GeV$ for $B_{\rm h}=0.001$ is higher
than the one $(4.6\cdot10^4-5.7\cdot10^5)~\GeV$ allowed for
$B_{\rm h}=1$. The same hierarchy holds for $\Ti$'s too. Finally,
we remark that the maximal $\vHi$'s depend very weakly on $\Ti$.

Concluding, we can say that although the $\Gr$ constraint is more
severe in the present QS than in the case of a pure KD era, it
remains much weaker than in the case of the SC. As a consequence,
relatively high values of $\Ti$ can be comfortably accommodated in
both former cases.


\end{document}